\documentclass[12pt]{article}
\usepackage{latexsym,amsthm,amssymb,amsfonts,amsbsy,graphicx,lscape,array,citesort,multibox}

\def\a{\alpha}
\def\b{\beta}
\def\g{\gamma}
\def\c{\gamma}
\def\d{\delta}
\def\e{\eta}
\def\h{\eta}
\def\l{\lambda}

\def\m{\mu}
\def\n{\nu}
\def\r{\rho}
\def\o{\omega}

\def\s{\sigma}

\def\t{\tau}

\def\e{\varepsilon}

\def\pa{\partial}

\def\ll{\left}
\def\rr{\right}

\def\be{\begin{equation}}
\def\ee{\end{equation}}
\def\beq{\begin{eqnarray}}
\def\eeq{\end{eqnarray}}
\def\nn{\nonumber}

\def\cg{{\cal G}}

\def\co{{\cal O}}

\def\cw{{\cal W}}

\def\cy{{\cal Y}}

\newcommand{\bqn}{\begin{eqnarray}}\newcommand{\eqn}{\end{eqnarray}}
\newtheorem{theorem}{Theorem}
\newtheorem{lemma}{Lemma}

\newtheorem{prop}{Proposition}

\usepackage{color}
\definecolor{rougef}{rgb}{0.56,0,0}
\definecolor{vertf}{rgb}{0,0.5,0}
\definecolor{bleuf}{rgb}{0,0,0.8}

\newcommand{\TBC}[1]{\textcolor{rougef}{\textbf{[TO BE CHECKED!!]}}}
\begin{document}
\begin{flushright}
IHES/P/05/25 \\
ULB-TH/05-18
\end{flushright}

\vspace{.6cm}

\begin{centering}

{\Large {\bf Spin three gauge theory revisited}}


\vspace{.2cm}

\begin{center}
{Xavier Bekaert$^{a,}$\footnote{E-mail address:
\tt{bekaert@ihes.fr}}, Nicolas Boulanger$^{b,}$\footnote{Charg\'e
de Recherches FNRS (Belgium); \tt{nicolas.boulanger@umh.ac.be}}
and Sandrine Cnockaert$^{c,}$\footnote{Aspirant du FNRS (Belgium);
\tt{sandrine.cnockaert@ulb.ac.be}} }
\end{center}

{\small{
\begin{center}
$^a$
Institut des Hautes \'Etudes Scientifiques\\
Le Bois-Marie, 35 route de Chartres, 91440 Bures-sur-Yvette (France)\\
$^b$
Universit\'e de Mons-Hainaut,
M\'ecanique et Gravitation
\\ 6 avenue du Champ de Mars, 7000 Mons (Belgium)\\
$^c$
Physique Th\'eorique et Math\'ematique, Universit\'e Libre
de Bruxelles\\
and International Solvay Institutes,\\
U.L.B. Campus Plaine, C.P. 231, B-1050, Bruxelles (Belgium)\\
\end{center}}}
\end{centering}

\vspace*{.2cm}

\begin{abstract}
We study the problem of consistent interactions for spin-$3$ gauge
fields in flat spacetime of arbitrary dimension $n > 3$.
Under the sole assumptions of Poincar\'e and parity
invariance, local and perturbative deformation of the free
theory, we determine all nontrivial consistent deformations of
the abelian gauge algebra and classify the corresponding
deformations of the quadratic action, at first order in the
deformation parameter. We prove that all such vertices are cubic,
contain a total of either three or five derivatives and are
uniquely characterized by a rank-three constant tensor (an internal 
algebra structure constant). The covariant cubic vertex containing
three derivatives is the vertex discovered by Berends, Burgers and
van Dam, which however leads to inconsistencies at second order in
the deformation parameter. In dimensions $n>4$ and for a
completely antisymmetric structure constant tensor, another covariant cubic vertex
exists, which contains five derivatives and passes the consistency test
where the previous vertex failed. \end{abstract}

\vspace*{3cm}

\pagebreak

\section{Introduction}
\label{sec:Introduction}

Whereas gauge theories describing free massless fields of
arbitrary high spin are by now well established, it still remains
unclear whether nontrivial consistent self-couplings and/or
cross-couplings among those fields may exist at the level of the
action, such that the deformed gauge algebra is non-abelian. The
old Fronsdal programme of introducing consistent couplings among
higher-spin gauge fields \cite{Fronsdal:1978rb} is still far away
from completion. Actually, there is a general belief that such
interactions are forbidden, except perhaps when the cosmological
constant is nonvanishing, in which case encouraging results have
been found at the level of equations of motion (see e.g.
\cite{Vasiliev:2004qz,Sagnotti:2005} and references therein).

The Fronsdal programme was initially investigated in two distinct
directions: either searching for consistent vertices for
higher-spin gauge fields interacting with each other but not with
gravity, or attempting to couple consistently some given
higher-spin gauge field with gravity. On the one hand, the problem
of consistent interactions among higher-spin gauge fields in
Minkowski spacetime ${\mathbb R}^{n-1,1}$ was addressed in
\cite{deWit:1979pe,Bengtsson:1983pd,Berends:1984wp,Bengtsson:1983bp,Bengtsson:1985iw,Berends:1984rq,Berends:1985xx,Bengtsson:1986bz,Damour:1987fp,Bengtsson:1986kh,Bengtsson:1987jt,Fradkin:1991iy}
where some positive results have been obtained. In the light-cone
gauge, three-point couplings between completely
symmetric\footnote{Light-cone cubic vertices involving mixed
symmetry gauge fields were computed in dimensions $n=5,6$
\cite{Metsaev}.} gauge fields with arbitrary spins $s>2$, were
constructed in \cite{Bengtsson:1983pd,Bengtsson:1986kh,Fradkin:1991iy}. 
For the pure spin-$3$ case, a cubic vertex was obtained in a covariant
form by Berends, Burgers and van Dam \cite{Berends:1984wp}. These
results describe consistent interactions at first order in a
deformation parameter $g$ and involve higher-derivatives. However,
no-go results soon demonstrated the impossibility of extending
these interactions to the next orders in powers of $g$ for the
pure spin-$3$ case
\cite{Bengtsson:1983bp,Berends:1984rq,Bengtsson:1986bz}. On the
other hand, the first explicit attempts to introduce interactions
between higher-spin gauge fields and gravity encountered severe
problems \cite{diff}.

Very early, the idea was proposed that a consistent higher-spin
gauge theory could exist, provided all spins are taken into
account \cite{Fronsdal:1978rb}. In order to overcome the
gravitational coupling problem, it was also suggested to perturb
around a curved, conformally-flat background, like for example $AdS_n$. 
In such a case, the cosmological constant $\Lambda$ can be used to cancel
the positive mass dimensions appearing with the increasingly many
derivatives of the vertices. As the works of Fradkin, Vasiliev and
others show, interesting results have indeed been obtained in
those directions, even at the level of the action
\cite{Fradkin:1987ks}.

If there is a lesson to learn from decades of efforts toward a
consistent theory of interacting higher-spin gauge fields, it
certainly is the unusual character of the possible interactions.
For instance, the cubic vertices contain more than two
derivatives.\footnote{The full theory presented in \cite{Vasiliev:2004qz}
is even expected to be non-local.} This, in turn, can be linked to
the fact that the spin-$s$ curvature is expressed via $s$
derivatives of the gauge field
\cite{Weinberg:1965rz,deWit:1979pe}. Consequently, in order to
investigate further the possible local higher-spin consistent
interactions, it is of prime importance to use as general a tool
as possible. A cohomological method is known
\cite{Barnich:1993vg}, which offers all the generality one could
wish and clearly organizes the calculation of the nontrivial
consistent couplings. In this approach, the old Noether method
(see for instance \cite{Berends:1984rq}) is reformulated in the
BRST framework where consistent couplings define deformations of
the solution of the master equation. This formulation has been
used recently in different contexts (see e.g.
\cite{Henneaux:1997bm,Boulanger:2000rq,Bekaert:2004dz} and
references therein).

In the present paper, we come back to the  initial (and more
modest) problem of consistent interactions among higher-spin gauge
fields in flat spacetime and concentrate on the pure spin-3 case.
The motivation behind our work is the existence of the new method 
\cite{Barnich:1993vg} developed in the meantime, which allows  
for an exhaustive treatment of the consistent interaction problem while, 
in the aforementioned works
\cite{Bengtsson:1983pd,Berends:1984wp,Bengtsson:1983bp,Bengtsson:1985iw,Berends:1984rq,Berends:1985xx,Bengtsson:1986bz,Bengtsson:1986kh,Bengtsson:1987jt,Fradkin:1991iy},
classes of deformation candidates were rejected {\textit{ab
initio} from the analysis for the sake of simplicity. For example,
spin-$3$ cubic vertices containing more than $3$ derivatives were not
considered in the otherwise very general analysis of
\cite{Berends:1984wp}. This ansatz was too restrictive since another
cubic vertex with five derivatives exists in dimensions higher than four (it is written explicitly in Appendix \ref{azerodeux}). 
Moreover, without fixing {\textit{a priori}} the maximal number of derivatives, 
we show that vertices deforming the gauge algebra must contain a total number of either
three or five derivatives.\footnote{This result is in agreement
with the general upper bound $k<s_1+s_2+s_3$ on the total number
$k$ of derivatives in a cubic vertex containing completely
symmetric fields of respective spin $s_1$, $s_2$ and $s_3$ \cite{Fradkin:1991iy}.}

The paper is organized as follows. In Section
\ref{sec:FreeTheory}, we review the free theory of massless spin-$3$
gauge fields represented by completely symmetric rank-$3$ tensors.
Our principal hypotheses are spelled out in Section \ref{sec:hypotheses}
and our main results are collected in Theorems \ref{galgebradefs}
and \ref{cvertices} presented in Section \ref{sec:mres}. The
section \ref{sec:BRSTSettings} gathers together the main BRST
results needed for the exhaustive treatment of the interaction problem:
The BRST spectrum of the theory is presented in Section
\ref{BRSTspectrum}. Some cohomological results have already been
obtained in \cite{Bekaert:2005ka}, such as the cohomology $H^*(\g)$
of the gauge differential $\gamma$ and the so called
characteristic cohomology $H_k^n(\d\vert d)$ in antighost number
$k\geqslant 2$. We recall the content of these groups in Sections
\ref{cohogamma} and \ref{Characteristiccohomology}. The
calculation of the invariant characteristic cohomology $H_k^{n}(\d
\vert d,H(\g))$ constitutes the core of the BRST analysis and is
achieved in Section \ref{Invariantcharacteristiccohomology}. The
self-interaction question is answered in Section
\ref{interactions}. We give our conclusions and discuss several
directions for future research in Section \ref{conclusions}.

\section{Free theory}
\label{sec:FreeTheory}

The local action for a collection
$\{h^a_{\m\n\r}\}$ of $N$ non-interacting completely symmetric
massless spin-3 gauge fields in flat spacetime is
\cite{Fronsdal:1978rb}
\begin{eqnarray}
S_0[h^a_{\m\n\r}] = \sum_{a=1}^N \int d^n x
                &\big[& -\frac{1}{2}\,\pa_{\s}h^a_{\m\n\r}\pa^{\s}h^{a\m\n\r} +
                        \frac{3}{2}\,\pa^{\m}h^a_{\m\r\s}\pa_{\n}h^{a\n\r\s} +
          \nonumber \\
               && \frac{3}{2}\,\pa_{\m}h^a_{\n}\pa^{\m}h^{a\n} +
                  \frac{3}{4}\,\pa_{\m}h^{a\m}\pa_{\n}h^{a\n} -
                        3 \,\pa_{\m}h^a_{\n}\pa_{\r}h^{a\r\m\n} \,\,\,\big]\,,
\label{freeaction}
\end{eqnarray}
where $h^a_{\m}:=\eta^{\n\r}h^a_{\m\n\r}\,$.
The Latin indices are internal indices taking $N$ values. They are raised
and lowered with the Kronecker delta's $\d^{ab}$ and $\d_{ab}$.
The Greek indices are space-time indices taking $n$ values, which are
lowered (resp. raised) with the ``mostly plus" Minkowski metric $\eta_{\m\n}$ (resp. $\eta^{\m\n}$).

The action (\ref{freeaction}) is invariant under the gauge
transformations
\begin{eqnarray}
        \delta_{\l}h^a_{\m\n\r} = 3 \,\pa^{}_{(\m}\l^a_{\n\r)}\,,\quad \eta^{\m\n}\l^a_{\m\n}\equiv 0\,,
\label{gaugetransfo}
\end{eqnarray}
where the gauge parameters $\l^a_{\n\r}$ are symmetric and traceless\footnote{Quadratic non-local actions \cite{FS} have been proposed in order to get rid of the trace 
constraint (\ref{gaugetransfo}) on the gauge parameter. Since
locality is an important hypothesis of the present work, we do not
discuss the non-local formulation here. Notice that by introducing a pure gauge field (sometimes refered to as ``compensator"), it is possible to write a local (but higher-derivative) action for spin-$3$ \cite{FS} that is invariant under 
unconstrained gauge transformations.
Very recently, this action was generalized to the arbitrary spin-$s$ case by further adding an auxiliary field \cite{Francia:2005bu} (see also \cite{Pashnev:1998ti} for an older ``non-minimal" version of it).}.
Curved (resp. square) brackets on spacetime indices denote strength-one complete symmetrization
(resp. antisymmetrization) of the indices.
The gauge transformations (\ref{gaugetransfo}) are abelian and irreducible.

The field equations read
\begin{eqnarray}
        \frac{\d S_0}{\d h^a_{\m\n\r}} \equiv G^{\m\n\r}_a = 0\,,
        \label{eom}
\end{eqnarray}
where
\begin{eqnarray}
G^{a}_{\m\n\r}:=F^{a}_{\m\n\r} -\frac{3}{2}
\eta^{}_{(\m\n}F^a_{\r)} \label{einstein}
\end{eqnarray}
is the ``Einstein" tensor and $F^{a}_{\m\n\r}$ the Fronsdal (or
``Ricci") tensor
\begin{eqnarray}
        F^{a}_{\m\n\r} :=
        \Box h^{a}_{\m\n\r} - 3 \,\pa^{\s}\pa^{}_{(\m}h^a_{\n\r)\s} + 3 \, \pa^{}_{(\m}\pa^{}_{\n}h_{\r)}^a\,.
\label{Fronsdal}
\end{eqnarray}
The Fronsdal tensor is gauge invariant thanks to the tracelessness
of the gauge parameters. Because we have $\d_{\l}S_0[h^a_{\m\n\r}]
= 0\,$ for the gauge transformations (\ref{gaugetransfo}), the
Einstein tensor $G^{a}_{\m\n\r}$ satisfies the Noether identities
\begin{eqnarray}
        \pa^{\r}G^{a}_{\m\n\r}-\frac{1}{n}\,\eta_{\m\n}\pa^{\r}G^{a}_{\r}\equiv 0
        \quad \quad (G^{a}_{\r} := \eta^{\m\n} G^{a}_{\m\n\r})
\label{Noether}
\end{eqnarray}
related to the symmetries of the gauge parameters $\l^a_{\m\n}\,$; in other words,
the l.h.s. of (\ref{Noether}) is symmetric and traceless.

The gauge symmetries enable one to get rid of some components of
$h^a_{\m\n\r}\,$, leaving it on-shell with $N^n_3$ independent
physical components, where $N^n_3$ is the dimension of the
irreducible representation of the ``little group" $O(n-2)$
($n\geqslant 3$) corresponding to a completely symmetric rank $3$
traceless tensor in dimension $n-2$. One has
$N^n_3=\frac{n^3-3n^2-4n+12}{6}\,$. Of course, $N^{4}_3=2$ for the
two helicity states $\pm 3\,$ in dimension $n=4\,$. Note also that
there is no propagating physical degree of freedom in $n=3$ since
$N^{3}_3=0$, so that we restrict our present work to $n>3$.  

An important object is the curvature (or ``Riemann") tensor
\cite{Weinberg:1965rz,deWit:1979pe,Damour:1987vm}
\begin{eqnarray}
        K^a_{\a\m|\b\n|\g\r}:= 8 \pa^{}_{[\g} \pa^{}_{[\b}\pa^{}_{[\a}h^a_{\m]\n]\r]}
\end{eqnarray}
which is antisymmetric in $\a\m\,$, $\b\n\,$, $\g\r$ and invariant under gauge transformations
(\ref{gaugetransfo}), where the gauge parameters $\lambda^a_{\m\n}$ are however {\textit{not}} necessarily traceless.

Its importance, apart from gauge invariance with unconstrained gauge parameters, stems from the fact that the field equations ({\ref{eom}}) are equivalent\footnote{As usual in field
 theory, we work in a space of smooth functions that vanish at infinity. 
 In particular, polynomials in $x^{\m}$ are forbidden.}
 to the following equations
\begin{eqnarray}
        \eta^{\a\b} K^a_{\a\m|\b\n|\g\r} = 0\,.
\end{eqnarray}
This was proved in the work \cite{BB} by combining various former
results \cite{Damour:1987vm,DVH,FS}.
%
\section{Deformations of the free theory}
\label{sec:defs}

\subsection{Basic assumptions}
\label{sec:hypotheses}

We assume, as in the traditional Noether deformation procedure,
that the deformed action can be expressed as a power series in a
coupling constant $g\,$, the zeroth-order term in the expansion
describing the free theory $S_0\,$:$$S=S_0+g\,S_1+\co(g^2)\,.$$
The procedure is then perturbative: one tries to construct the
deformations order by order in the deformation parameter $g\,$.

Some physical requirements naturally come out:
\begin{itemize}
  \item \underline{Poincar\'e and parity symmetry}: We ask that 
  the deformed Lagrangian be invariant under the
{\em{Poincar\'e}} group. Therefore, it should not depend explicitly
on the space-time cartesian coordinates $\{x^\mu\}$. The Lagrangian is moreover
required to be invariant under the parity transformation. This
implies that all Greek indices have to be contracted by means of
the Minkowski metric only.
  \item \underline{Nontriviality}: We reject {\em trivial} deformations
arising from field-redefinitions that reduce to the identity at
order $g^0\,$:
\begin{eqnarray}
\phi\longrightarrow \phi'=\phi+g\, \varphi (\phi, \pa \phi,
\cdots)+\co(g^2)\,. \label{fieldredef}
\end{eqnarray}
  \item \underline{Consistency}: A deformation of a
theory is called {\em{consistent}} if the deformed theory
possesses the same number of (possibly deformed) independent gauge
symmetries, reducibility identities, {\it etc.}, as the system we
started with. In other words, the number of physical degrees of
freedom is unchanged.
  \item \underline{Locality}: The deformed action $S[\phi]$ must be a {\em local}
  functional. The deformations of the gauge transformations, {\it etc.},
  must be local functions, as well as the allowed field redefinitions.
\end{itemize}
We remind the reader that a local function of some set of fields
$\phi^i$ is a smooth function of the fields $\phi^i$ and
their derivatives $\partial\phi^i$, $\partial^2\phi^i$, ...
up to some {\it finite} order, say $k$, in the number of
derivatives. Such a set of variables $\phi^i$,
$\partial\phi^i$, ..., $\partial^k\phi^i$ will be
collectively denoted by $[\phi^i]$. Therefore, a local function
of $\phi^i$ is denoted by $f([\phi^i])$. A local $p$-form
$(0\leqslant p \leqslant n)$ is a differential $p$-form the
components of which are local functions:
\begin{eqnarray}
        \omega =
\frac{1}{p!}\,\omega_{\m_1\ldots\m_p}(x, [\phi^i])\, dx^{\m_1}
\wedge \cdots \wedge dx^{\m_{p}}\,.\nonumber
\end{eqnarray}
A local functional is the integral of a local $n$-form.

\subsection{Main results}
\label{sec:mres}

Theorems \ref{galgebradefs} and \ref{cvertices} are presented in
this Section. They constitute strong yes-go and no-go theorems
that generalize previous works on spin-$3$ self-interactions.

\begin{theorem}\label{galgebradefs}Let $h^a_{\m\n\r}$ be
a collection of spin-$3$ gauge fields ($a=1,\ldots,N$) described by
the local and quadratic action of Fronsdal, in dimension $n>3$.

At first order in some smooth deformation parameter, the
nontrivial consistent local deformations of the (abelian) gauge
algebra that are invariant under parity and Poincar\'e
transformations, may always be assumed to be closed off-shell and are
in one-to-one correspondence with the structure constant tensors
$$C^a{}_{bc}=-C^a{}_{cb}$$ of an anticommutative internal algebra, that may be taken
as deformation parameters.

Moreover, the most general gauge transformations deforming the
gauge algebra at first order in $C=(f,g)$ are equal to
\begin{eqnarray}
        &\delta_{\l}h^a_{\m\n\r} &= 3\,\pa^{}_{(\m}\l^a_{\n\r)}+
        f^a{}_{bc}\,\Phi^{bc}_{\m\n\r}+\,g^a{}_{bc}\,(\Psi^{bc}_{\m\n\r}
        -{1\over n}\,\eta^{}_{(\m\n}\Psi^{bc}_{\r)})+\co(C^2)\,,
\label{defgtransfo}
\end{eqnarray}
up to gauge transformations
that either are trivial or do not deform the gauge algebra at
first order, where $\Phi^{bc}_{\m\n\r}$ and $\Psi^{bc}_{\m\n\r}$
are bilinear local functions of the gauge field
$h^a_{\m\n\r}$ and the traceless gauge parameter
$\l^a_{\m\n}$. 
The expression for $\Phi$ is lengthy and thus given in the appendix \ref{azeroun},
while 
\begin{eqnarray}
	\Psi^{bc}_{\m\n\r}=-{1\over 3}\,\eta^{\a\b}\partial^{}_{[\m}h^b_{\a]\n[\s,\t]}
	\partial^{}_{[\r}\l^{c\,\,\s,\t}_{\b]}+\mbox{perms}\,,
	\label{Psi}
\end{eqnarray}
where a coma denotes a partial derivative\footnote{For example
$\Phi^i_{,\,\a}\equiv\pa_{\a}\Phi^i$.} and ``perms" stands for the sum of terms obtained via all nontrivial permutations of the indices $\m\,,\n\,,\r\,$ from the first term of the r.h.s.

\noindent The structure constant tensors $f^a{}_{bc}$ and
$g^a{}_{bc}$ are some arbitrary constant tensors that are
antisymmetric in the indices $bc$. In mass units, the coupling constant $f^a{}_{bc}$ has dimension $-n/2$ and
$g^a_{bc}$ has dimension $-2-n/2$.

Both of these deformations exist in any dimension $n \geqslant 5$. 
In the case $n=4\,$, the structure constant tensor $g^a{}_{bc}$ vanishes. 
\end{theorem}
\noindent
Firstly, we found a deformation of the gauge symmetries 
(the one corresponding to the coefficients $g^a{}_{bc}$) which had not been 
written explicitly in previous spin-3 analyzes in flat space-time. Secondly,
without imposing any restriction on the maximal number of derivatives 
(as was implicit in most former works) we prove that the allowed possibilities are 
extremely restricted. 

An important question is whether these algebra deformations  
can be obtained from an appropriate flat space-time limit of the $(A)dS_n$ 
higher-spin algebras containing a finite-dimensional non-Abelian internal subalgebra
(studied in details by Vasiliev and collaborators \cite{Vasiliev:1986qx}).
An indication that this might be the case is provided by the 
deformation of the gauge transformations Eq. (\ref{defgtransfo}) involving 
the tensor $\Psi^{ab}_{\m\n\r}$. 
The presence of the term $\partial^{}_{[\m}h^b_{\a]\n[\s,\t]}$ in (\ref{Psi})
is reminiscent of the second frame-like connection (see e.g. the second reference 
of \cite{Sagnotti:2005}). They both involve two derivatives of the spin-3 field and
have the $gl(n)$-symmetry corresponding to the Young diagram 
{\footnotesize{
\begin{picture}(35,14)(0,0)
\multiframe(1,4)(10.5,0){3}(10,10){$$}{$$}{$ $}
\multiframe(1,-6.5)(10.5,0){2}(10,10){$$}{$$}
\end{picture}}}. 
More comments in that direction are given in sections \ref{cohogamma} and \ref{defdeux}. 

Another important physical question is whether or not these first-order
gauge symmetry deformations possess some Lagrangian counterpart,
\textit{i.e.} if there exist vertices that are invariant under
(\ref{defgtransfo}) at first order in $C$. The following theorem
provides a sufficient condition for that:
\begin{theorem}\label{cvertices}
Let  the constant tensor $C_{abc}=(f_{abc},g_{abc})$ be
completely antisymmetric, where $C_{abc}:=\d_{ad}C^d{}_{bc}\,$. Then,

$\bullet$ The quadratic local action (\ref{freeaction}) in dimension $n>3$ admits a
first-order consistent deformation \be
S[h^a_{\m\n\r}]\,=\,S_0\,+\,f_{abc}\,S^{abc}\,+\,g_{abc}\,T^{abc}\,+\,\co(C^2)\,,\label{defcubact}\ee
which is gauge invariant under the deformed gauge transformations
(\ref{defgtransfo}) at first order in the deformation parameters.
Furthermore, this antisymmetry condition on the tensor $f^a{}_{bc}$ is necessary for
the existence of the corresponding deformation of the action.

$\bullet$ The vertices in the first-order
deformations are determined uniquely by the structure constants $f_{abc}$ and
$g_{abc}$, modulo vertices that do not
deform the gauge algebra. The corresponding local functionals
$S^{abc}[h^d_{\m\n\r}]$ and $T^{abc}[h^d_{\m\n\r}]$ are cubic in
the gauge field and respectively contain three and five
derivatives. Actually, there are no other nontrivial
consistent vertices containing at most three derivatives that
deform the gauge transformation at first order.

$\bullet$ At second order in $C$, the deformation of the
gauge algebra can be assumed to close off-shell without loss of generality,
but it is obstructed if and only if $f_{abc}\neq 0\,$.  
\end{theorem}

\noindent The first-order covariant cubic deformation
$S^{bc}{}_a[h^d_{\m\n\r}]$ is the Berends--Burgers--van Dam vertex
\cite{Berends:1984wp} (reviewed for completeness in Appendix
\ref{azeroun}) while the other cubic deformation
$T^{bc}{}_a[h^d_{\m\n\r}]$ is written in Appendix
\ref{azerodeux}. We do not know yet if the antisymmetry condition
on the structure constant $g^a_{bc}$ is necessary or not for the
existence of a consistent vertex at first order.

It is possible to provide a more intrinsic characterization of the
conditions on the constant tensors. Let $\cal A$
be an {\it anticommutative} algebra of dimension $N$ with a basis
$\{T_a\}$ . Its multiplication law $\ast:{\cal A}^2\rightarrow \cal
A$ obeys $a\ast b=-b\ast a$ for any $a,b\in\cal A$, which is equivalent to
the fact that the structure constant tensor $C^a{}_{bc}$ defined
by $T_b\ast T_c=C^a{}_{bc}\,T_a$ is antisymmetric in the covariant
indices: $C^a{}_{bc}=-C^a{}_{cb}$. Moreover, let us assume that the
algebra $\cal A$ is a Euclidean space, {\it i.e.} it is endowed
with a scalar product $\langle\,\,,\,\rangle:{\cal A}^2\rightarrow
\mathbb R$ with respect to which the basis $\{T_a\}$ is
orthonormal, $\langle\,T_a\,,\,T_b\,\rangle=\d_{ab}$.
For an anticommutative algebra, the scalar
product is said to be {\it invariant} (under the left or right
multiplication) if and only if
$\langle\,a\ast b\,,\,c\,\rangle=\langle\,a\,,\,b\ast c\,\rangle$ for any
$a,b,c\in \cal A\,$, and the latter
property is equivalent to the complete antisymmetry of the
trilinear form
$$C:{\cal A}^3\rightarrow{\mathbb R}:(a,b,c)\mapsto
C(a,b,c)=\langle\,a\,,\,b\ast c\,\rangle$$ or, in
components, to the complete antisymmetry property of the covariant
tensor $C_{abc}:=\d_{ad}\,C^d{}_{bc}$.

The gauge algebra
inferred from the Berends--Burgers--van Dam vertex is inconsistent
at second order \cite{Bengtsson:1983bp,Berends:1984rq} and no
corresponding quartic interaction can be constructed
\cite{Bengtsson:1986bz}.
Originally, consistency of the Berends--Burgers--van Dam deformation at second order was shown to require that $f^d{}_{ec}f^e{}_{ab}=f^d{}_{ae}f^e{}_{bc}$ \cite{Berends:1984rq}, which means that the corresponding internal algebra is associative $(a\ast b)\ast c=a\ast (b\ast c)$.
In Section \ref{obstr}, we actually obtain a stronger condition from consistency: $f^d{}_{ec}f^e{}_{ab}=0$, {\it i.e.} the internal algebra is nilpotent of order three: $(a\ast b)\ast c=0$. In any case, to derive that the Berends--Burgers--van Dam vertex is inconsistent at order two, one may use the following well-known lemma
\begin{lemma}\label{lemalg}
If an anticommutative algebra endowed with an invariant
scalar product is associative, then the product of any two elements is zero (in other words, the algebra is nilpotent of order two).
\end{lemma}
\noindent{\bfseries{Proof}}: Under the hypotheses of Lemma \ref{lemalg}, one gets  $\langle\,a\ast b\,,\,b\ast a\,\rangle=\langle\,a\,,\,b\ast (b\ast a)\,\rangle=\langle\,a\,,\,(b\ast b)\ast a\,\rangle=0$ which implies $a\ast b=0$ for any $a,b\in \cal A$. \qed

An exciting result is that the second
deformation corresponding to $g_{abc}=g_{[abc]}$ passes the gauge
{\em{algebra}} consistency requirement where the vertex of
Berends, Burgers and van Dam fails. Unfortunately, we do not know
if there exist second order gauge {\em{transformations}} that are
consistent at this order.

The proofs of Theorems \ref{galgebradefs} and \ref{cvertices} are
given in Section \ref{interactions}. They rely on a BRST
cohomological reformulation presented in the next Section.
%
%
\section{BRST settings}
\label{sec:BRSTSettings}
%
\subsection{BRST spectrum and differential}
\label{BRSTspectrum}
%
According to the general rules of the BRST-antifield formalism, a grassmann-odd ghost
$C_{\m\n}^a$ is introduced, which accompanies each grassmann-even gauge parameter $\l_{\m\n}^a$.
In particular, it possesses the same algebraic symmetries as $\l_{\m\n}^a$: it is symmetric and
traceless in its spacetime indices.
Then, to each field and ghost of the spectrum, a corresponding
antifield (or antighost) is added, with the same algebraic symmetries but
the opposite Grassmann parity. A $\mathbb{Z}$-grading called {\textit{ghost number}} ($gh$) is associated with the BRST differential $s$, while the {\textit{antighost number}} ($antigh$)
of the antifield $Z^*$ associated with the field (or ghost) $Z$
is given by $antigh(Z^*)\equiv gh(Z)+1\,$.
More precisely, in the theory under consideration, the spectrum of fields (including ghosts)
and antifields together with their respective ghost and antighost numbers is given by
\begin{itemize}
\item the fields $h^a_{\m\n\r}\,$, with ghost number $0$ and antighost number $0$;
\item the ghosts $C^a_{\m\n}$, with ghost number $1$ and antighost number $0$;
\item the antifields $h^{*\m\n\r}_a$, with ghost number $-1$ and antighost number $1$;
\item the antifields $C^{*\m\n}_a$, with ghost number $-2$ and antighost number $2\,$.
\end{itemize}
The BRST differential $s$ of  the free theory (\ref{freeaction}), (\ref{gaugetransfo}) is generated by the functional
\begin{eqnarray}
W_0 = S_0 [h^a] \;+ \int d^nx\; ( 3\, h^{*\m\n\r}_a \, \pa_{\m} C_{\n\r}^a )\,. \nonumber
\end{eqnarray}
More precisely, $W_0$ is the generator of the BRST differential $s$ of the free theory through
\begin{eqnarray}
        s A = (W_0, A)_{a.b.}\,, \nonumber
\end{eqnarray}
where the antibracket $(~,~)_{a.b.}$ is defined by
\begin{eqnarray}
(A,B)_{a.b.}=\frac{\d^R A}{\d \Phi^I}\frac{\d^L B}{\d \Phi^*_I} -
 \frac{\d^R A}{\d \Phi^*_I}\frac{\d^L B}{\d \Phi^I}\,.
\end{eqnarray}
The functional $W_0$ is a solution of the \emph{master equation}
\begin{eqnarray}
        (W_0,W_0)_{a.b.}=0\,.
\end{eqnarray}
In the theory at hand, the BRST-differential $s$ decomposes into $s=\g + \d \,$.
The first piece $\g\,$, the differential along the gauge orbits, is associated with another
grading called \textit{pureghost number} ($puregh$) and increases it by one unit, whereas the Koszul-Tate differential $\d$ decreases the antighost (or antifield) number by one unit.
The differential $s$ increases the ghost number by one unit.
Furthermore, the ghost, antighost and pureghost gradings are not
independent. We have the relation
\begin{eqnarray}
        gh = puregh - antigh \,.
\end{eqnarray}

The pureghost number, antighost number, ghost number and grassmannian parity of the various fields are displayed in Table \ref{table1}.

\begin{table}[!ht]
\centering
\begin{tabular}{|c|c|c|c|c|}
\hline Z  & $puregh(Z)$  & $antigh(Z)$  & $gh(Z)$  & parity (mod $2$)\\ \hline
$h^a_{\m\n\r}$   &$0$  & $0$  &$0$ &$0$ \\
$C^a_{\m\n}$ & $1$ & $0$ & $1$ & $1$ \\
$h^{*\m\n\r}_a$ & $0$& $1$ & $-1$ & $1$ \\
$C^{*\m\n}_a$ & $0$ & $2$ & $-2$ & $0$ \\
\hline
\end{tabular}
\caption{\it pureghost number, antighost number, ghost number
and parity of the (anti)fields.\label{table1}}
\end{table}
The action of the differentials  $\delta$ and $\gamma$ gives zero on all the
fields of the formalism except in the few following cases:
\begin{eqnarray}
\d h^{*\m\n\r}_a &=&G^{\m\n\r}_a\,,
\nonumber \\
\d C^{*\m\n}_a &=& -3 ( \pa_{\r}h^{*\m\n\r}_a - \frac{1}{n}\eta^{\m\n}\pa_{\r}h^{*\r}_a )\,,
\nonumber \\
\gamma h^a_{\m\n\r} &=& 3\,\pa^{}_{(\m}C_{\n\r)}^a\,.
\nonumber \\
\end{eqnarray}
%
\subsection{BRST deformation}
\label{deformation}
%
As shown in \cite{Barnich:1993vg}, the Noether procedure can be
reformulated within a BRST-cohomological framework. Any
consistent deformation of the gauge theory corresponds to a
solution $$W=W_0+g W_1+g^2W_2+\co(g^3)$$ of the deformed master
equation $(W,W)_{a.b.}=0$. Consequently, the first-order nontrivial
consistent local deformations $W_1=\int a^{n,\,0}$ are in
one-to-one correspondence with elements of the cohomology
$H^{n,\,0}(s \vert\, d)$ of the zeroth order BRST differential
$s=(W_0,\cdot)$ modulo the total derivative $d\,$, in maximum
form-degree $n$ and in ghost number $0\,$. That is, one must
compute the general solution of the cocycle condition
\begin{eqnarray}
        s a^{n,\,0} + db^{n-1,1} =0\,,
        \label{coc}
\end{eqnarray}
where $a^{n,\,0}$ is a top-form of ghost number zero and
$b^{n-1,1}$ a $(n-1)$-form of ghost number one, with the
understanding that two solutions of (\ref{coc}) that differ by a
trivial solution should be identified
\begin{eqnarray}
        a^{n,\,0}\sim a^{n,\,0} + s p^{n,-1}  + dq^{n-1,\,0} \nonumber
\end{eqnarray}
as they define the same interactions up to field redefinitions (\ref{fieldredef}).
The cocycles and coboundaries $a,b,p,q,\ldots\,$ are local forms of
the field variables (including ghosts and antifields).

The corresponding second-order interactions $W_2$ must satisfy the consistency condition
$$s  W_2=-\frac{1}{2} (W_1,W_1)_{a.b.}\,.$$ This condition is controlled by the local BRST cohomology group $H^{n,1}(s\vert d)$.

%
\subsection{Cohomology of $\g$}
\label{cohogamma}
%
In the context of local free theories in Minkowski space for
massless spin-$s$ gauge fields represented by completely symmetric
(and double traceless when $s>3$) rank $s$ tensors, the groups
$H^*(\g)$ have recently been calculated \cite{Bekaert:2005ka}.
Accordingly, we only recall the latter results in the special case
$s=3$ and introduce some new notations.
\begin{prop}\label{Hgamma} The cohomology of $\g$ is
isomorphic to the space of functions depending on
\begin{itemize}
  \item the antifields $h^{*\m\n\r}_a$, $C^{*\m\n}_a$ and their derivatives, denoted by
  $[\Phi^{*i}]\,$,
  \item the curvature and its derivatives $[K^a_{\a\m|\b\n|\g\r}]\,$,
  \item the symmetrized derivatives $\pa^{}_{(\a_1}\ldots\pa^{}_{\a_k}F^a_{\m\n\r)}$ of the Fronsdal tensor,
  \item the ghosts $C_{\m\n}^a$ and the traceless parts of $\pa^{}_{[\a}C_{\m]\n}^a$ and
  $\pa^{}_{[\a}C_{\m][\n,\b]}^a$.
\end{itemize} 
Thus, identifying with zero any $\gamma$-exact term in $H(\gamma)$, we have   
$$ \g f=0 $$
if and only if $$f=
f\left([\Phi^{*i}],[K^a_{\a\m|\b\n|\g\r}],\{F^a_{\m\n\r}\},
                               C_{\m\n}^a, \widehat{T}^a_{\a\m\vert\n}, \widehat{U}^a_{\a\m\vert\b\n}
     \right)$$
where $\{F^a_{\m\n\r}\}$ stands for the completely symmetrized
derivatives $\pa^{}_{(\a_1}\ldots\pa^{}_{\a_k}F^a_{\m\n\r)}$ of
the Fronsdal tensor, while $\widehat{T}^a_{\a\m\vert\n}$ denotes
the traceless part of $T^a_{\a\m\vert\n}:=\pa^{}_{[\a}C_{\m]\n}^a$ and $\widehat{U}^a_{\a\m\vert\b\n}$ the
traceless part of ${U}^a_{\a\m\vert\b\n}:=\pa^{}_{[\a}C_{\m][\n,\b]}^a\,$.
\end{prop}

This proposition provides the possibility of writing down the
most general gauge-invariant interaction terms. Such
higher-derivative Born-Infeld-like Lagrangians were already
considered in Ref. \cite{Damour:1987fp}. These deformations are
consistent to all orders but they do not deform the gauge
transformations (\ref{gaugetransfo}). Also notice that any
function of the Fronsdal tensor or its derivatives corresponds to
a field redefinition. \vspace{.3cm}

Let $\{\o^I\}$ be a basis of the space of polynomials in the
$C_{\m\n}^a$, $\widehat{T}^a_{\a\m\vert\n}$ and $\widehat{U}^a_{\a\m\vert\b\n}$
(since these variables anticommute, this space is finite-dimensional).
If a local form $a$ is $\gamma$-closed, we have
\begin{eqnarray}
        \g a = 0 \quad\Rightarrow\quad a \,=\,
        \a_J([\Phi^{i*}],[K],\{F\})\,
        \o^J(C_{\m\n}^a,\widehat{T}^a_{\a\m\vert\n},\widehat{U}^a_{\a\m\vert\b\n}) + \g b\,,
\end{eqnarray}
If $a$ has a fixed, finite ghost number, then $a$ can only contain
a finite number of antifields. Moreover, since the
{\textit{local}} form $a$ possesses a finite number of
derivatives, we find that the $\a_J$ are polynomials. Such a
polynomial $\a_J([\Phi^{i*}],[K],\{F\})$ will be called  an
{\textit{invariant polynomial}}.\vspace{.3cm}

\noindent {\textbf{Remark 1}:} Because of the Damour-Deser identity
\cite{Damour:1987vm}
$$\eta^{\a\b}K_{\a\m|\b\n|\g\r}\equiv 2\, \pa_{[\g}F_{\r]\m\n}\,,$$ the derivatives of the
Fronsdal tensor are not all independent  of the curvature tensor
$K$. This is why, in Proposition \ref{Hgamma}, the completely
symmetrized derivatives of $F$ appear, together with all the
derivatives of the curvature $K$. However, from now on, we will
assume that every time the trace $\eta^{\a\b}K_{\a\m|\b\n|\g\r}$
appears, we substitute $2\pa_{[\g}F_{\r]\m\n}$ for it. With this convention, we can write
$\a_J([\Phi^{i*}],[K],[F])$ instead of the unconvenient notation
$\a_J([\Phi^{i*}],[K],\{F\})$.\vspace{.3cm}

\noindent {\textbf{Remark 2}:} It is possible to make a link with the variables 
occurring in the 
frame-like first-order formulation of free massless spin-3 field in Minkowski space-time 
\cite{Vasiliev:1980as}. There, the spin-3 field is represented off-shell by a frame-like
object $e_{\m|ab}$, symmetric and traceless in the internal indices $(a,b)$. The spin-3
connection $\o_{\m|b|a_1a_2}$ is traceless in the internal Latin indices, 
symmetric in ($a_1,a_2$) and obeying $\o_{\m|(b|a_1a_2)}\equiv 0$. 
The gauge transformations are $\d e_{\m|ab} = \pa_{\m} \xi_{ab} + \a_{\m|ab}$, 
$\d \o_{\m|b|a_1a_2} = \pa_{\m}\a_{b|a_1a_2} + \Sigma_{\m|b|a_1a_2}$, where the parameter
$\xi_{ab}$ is symmetric and traceless in $(a,b)$, the generalized Lorentz parameter
$\a_{\m|ab}$ is completely traceless, symmetric in ($a,b$) and satisfies the identity 
$\a_{(\m|ab)}\equiv 0$. Finally, the parameter $\Sigma_{\m|a|bc}$ transforms in 
the $o(n-1,1)$ irreducible representation associated with the Young tableau 
{\footnotesize{
\begin{picture}(25,14)(0,0)
\multiframe(1,4)(10.5,0){2}(10,10){$\m$}{$a$}
\multiframe(1,-6.5)(10.5,0){2}(10,10){$b$}{$c$}
\end{picture}}},     
in the manifestly symmetric convention. 
By choosing the generalized Lorentz parameter appropriately, it is possible
to work in the gauge where the frame-field $e_{\m|ab}$ is completely symmetric, 
$e_{\m|ab}=e_{(\m|ab)}\equiv h_{\m a b}$. Then, it is still possible to perform a 
gauge transformation with parameters $\a_{\m|ab}$ and $\xi_{ab}$, 
provided the traceless component of $\pa_{[\m}\xi_{a]b}$ be equal to $-\a_{[\m|a]b}$. 
The traceless component of $\pa_{[\m}\xi_{a]b}$ is nothing but the variable 
$\widehat{T}_{\m\a\vert\b}$ in the BRST conventions. 
Furthermore, in the 1.5 formalism where the connection is still present in the action, 
but viewed as a function of $e_{\m|a_1a_2}$, consistency with the ``symmetric gauge" 
$e_{\m|ab}=e_{(\m|ab)}\equiv h_{\m a b}$ implies that the traceless component of 
the second derivative $\pa_{[a}\xi_{b][c,\mu]}$ be entirely determined by 
$\Sigma_{\m|b|ac}$. The traceless component of $\pa_{[a}\xi_{b][c,\mu]}$ is the 
variable $\widehat{U}_{\a\b\vert\c\m}$ in the BRST language.  
The relations $\widehat{T}_{\m\a\vert\b}\longleftrightarrow \a_{\m|ab}$
and $\widehat{U}_{\a\b\vert\c\m}\longleftrightarrow \Sigma_{\m|b|ac}$
are now manifest (note the we work in the manifestly antisymmetric convention, as 
opposed to the choice made in \cite{Vasiliev:1980as}). The variables 
$\{C_{\m\n},\widehat{T}_{\m\a\vert\b},\widehat{U}_{\a\b\vert\c\m}\}\in H(\gamma)$ 
in the ghost sector are in one-to-one correspondence with the gauge parameters 
$\{\xi_{\m\n},\a_{\m|ab},\Sigma_{\m|b|ac}\}$ of the first-order formalism 
\cite{Vasiliev:1980as}.   

\subsection{Invariant Poincar\'e lemma}
\label{invPlemma}

We shall need several standard results on the cohomology of $d$ in
the space of invariant polynomials.
\begin{prop}\label{2.2}
In form degree less than $n$ and in antifield number strictly greater than $0$,
the cohomology of $d$ is trivial in the space of invariant
polynomials.
That is to say, if $\a$ is an invariant polynomial, the equation
$d \a = 0$ with $antigh(\a) > 0$ implies
$ \a = d \b$ where $\b$ is also an invariant polynomial.
\end{prop}
\noindent The latter property is rather generic for gauge theories
(see e.g. Ref. \cite{Boulanger:2000rq} for a proof), as well as
the following:

\begin{prop}\label{csq}
If $a$ has strictly positive antifield number, then the equation
$\gamma a + d b = 0$ is equivalent, up to trivial redefinitions,
to $\gamma a = 0$. More precisely, one can always add $d$-exact
terms to $a$ and get a cocycle $a' := a  + d c$ of $\gamma$, such
that $\g a'= 0$.
\end{prop}
\vspace*{.2cm}

\noindent{\bfseries{Proof}}: Along the lines of Ref.
\cite{Boulanger:2000rq}, we consider the descent associated with
$\gamma a + d b = 0$: from this equation, one infers, by using the
properties $\gamma^2 = 0$, $\gamma d + d \gamma = 0$ and the
triviality of the cohomology of $d$, that $\gamma b + dc = 0$ for
some $c$.  Going on in the same way, we build a ``descent"
\begin{eqnarray}
\gamma a + d b &=& 0\nn\\\gamma b + dc &=& 0\nn\\\gamma c + de &=& 0\,,\nn\\ &\vdots&\label{descent}\\
\gamma m + dn &=& 0\,,\nn\\ \gamma n &=& 0\,.\nn
\end{eqnarray}
in which each successive equation has one less unit of
form-degree. The descent ends with $\gamma n = 0$ either because
$n$ is a zero-form, or because one stops earlier with a
$\gamma$-closed term. Now, because $n$ is $\gamma$-closed, one
has, up to trivial, irrelevant terms, $n = \a_J \omega^J$.
Inserting this into the previous equation in the descent yields
\be 
d (\a_J) \omega^J \pm \a_J d \omega^J + \gamma m = 0 .
\label{keya3} 
\ee 
In order to analyse this equation, we introduce a new differential.
\vspace{2mm}

\noindent \textbf{Definition (differential $D$)}: The action of the differential $D$ on
$h^a_{\mu \nu\rho}$, $h^{*\mu \nu\r}_a$, $C^{*\m\n}_a$ and all
their derivatives is the same as the action of the total
derivative $d$, but its action on the ghosts is given by :
\begin{eqnarray}
D C^a_{\m\n} &=& {\frac {4}{3}} \, d x^{\a}\, {\widehat{T}}^a_{\a(\mu\vert\nu )}\,,
\nonumber \\
D  {\widehat{T}}^a_{\m\a\vert\b} &=& d x^{\r} \, {\widehat{U}}^a_{\m\a\vert\r\b}\,,
\nonumber \\
D(\partial _{\rho _1 \ldots \rho _t} C_{\mu}) &=& 0 ~ {\rm \ if \
}~ t\geqslant 2 .
\end{eqnarray}
The above definitions follow from
\begin{eqnarray}
        \pa_{\a}C^a_{\m\n} &=& \frac{1}{3}(\gamma h^a_{\a\m\n})+\frac{4}{3}}T^a_{\a(\m\vert\n)\,,
        \nonumber \\
        \pa_{\r}T_{\m\a\vert\b} &=& -\frac{1}{2}\,\g(\pa_{[\a}h_{\m]\b\r})+U_{\m\a\vert\r\b}\,,
        \nonumber \\
        \pa_{\r}U_{\m\a\vert\n\b} &=& \frac{1}{3} \g (\pa_{[\m}h_{\a]\r[\b,\n]})\,.
\end{eqnarray}
The operator $D$ thus coincides with $d$ up to $\gamma$-exact terms.

It follows from the definitions that $D\omega^J = A^J{}_I
\omega^I$ for some constant matrix $A^J{}_I$ that involves $dx^\m$
only. One can rewrite (\ref{keya3}) as \be \underbrace{d
(\a_J) \omega^J \pm \a_J D \omega^J}_{=(d\a_J\,\pm\,\a_I
A^I{}_J)\omega^J} + \gamma m' = 0 \ee which implies, \be d (\a_J)
\omega^J \pm \a_J D \omega^J = 0 \label{keya4} \ee since a term of
the form $\b _J \omega^J $ (with $\b _J$ invariant) is $\g$-exact
if and only if it is zero. It is also convenient to introduce a
new grading.
\vspace{2mm}

\noindent\textbf{Definition ($D$-degree)}: The number of
${\widehat{T}}_{\a\m|\n}$'s plus two times the number of
${\widehat{U}}_{\a\m|\b\n}$'s is called the $D$-degree. It is
bounded because there is a finite number of
${\widehat{T}}_{\a\m|\n}$'s and ${\widehat{U}}_{\a\m|\b\n}$'s,
which are anticommuting.
 The operator $D$ splits as the
sum of an operator $D_1$ that raises the $D$-degree by one unit,
and an operator $D_0$ that leaves it unchanged. $D_0$ has the
same action as $d$ on $h_{\mu \nu\r}$, $h^{*\mu \nu\r}$,
$C^{*\alpha\b}$ and all their derivatives, and gives $0$ when
acting on the ghosts. $D_1$ gives $0$ when acting on all the
variables but the ghosts on which it reproduces the action of $D$.

Let us expand (\ref{keya3}) according to the $D$-degree. At lowest
order, we get \be d \a_{J_0} = 0 \ee where $J_0$ labels the
$\omega^J$ that contain no derivative of the ghosts ($D\omega^J
= D_1 \omega^J $ contains at least one derivative). This equation
implies, according to Proposition \ref{2.2}, that $\a_{J_0} = d \b
_{J_0}$ where $\b _{J_0}$ is an invariant polynomial. Accordingly,
one can write \be \a_{J_0}\omega^{J_0} = d(\b _{J_0} \omega^{J_0})
\mp \b _{J_0} D\omega^{J_0} + \hbox{ $\g$-exact terms}  . \ee The
term $\b _{J_0} D\omega^{J_0}$ has $D$-degree equal to $1$. Thus,
by adding trivial terms to the last term $n$($=\a_J\o^J$) in
the descent  (\ref{descent}), we can assume that it does not
contain any term of $D$-degree $0$. One can then successively
remove the terms of $D$-degree $1$, $D$-degree $2$, etc, until one
gets $n = 0$. One then repeats the argument for $m$ and the
previous terms in the descent (\ref{descent}) until one gets
$b = 0$, i.e., $\g a = 0$, as requested.\qed

\subsection{Cohomology of $\d$ modulo $d\,$: $H^n_k(\d \vert\, d)$}
\label{Characteristiccohomology}

In this section, we review the local Koszul-Tate cohomology
groups in top form-degree and antighost numbers $k\geqslant 2\,$.
The group $H^D_1(\d \vert\, d)$ describes the infinitely many
conserved currents and will not be studied here.
\vspace*{.2cm}

Let us first recall a general  theorem (Theorem 9.1 in \cite{Barnich:1994db}).

\begin{prop}\label{usefll}
For a linear gauge theory of reducibility order $r$,
\begin{eqnarray}
H_p^n(\d \vert\, d)=0\; for\; p>r+2\,. \nonumber
\end{eqnarray}
\end{prop}
Since the theory at hand has no reducibility, we are left with the computation of
$H_2^n(\d \vert\, d)\,$. The cohomology $H_2^n(\d \vert\, d)$ is given by the following
theorem.

\begin{prop}\label{H2}
A complete set of representatives of $H^n_2(\d\vert d)$ is given by the antifields
$C_a^{*\m\n}$, up to explicitly $x$-dependent terms. In detail,
\begin{eqnarray}
        \left.
        \begin{array}{ll}
         \delta a^n_2 + d b^{n-1}_1 = 0\,,
        \nonumber \\
         \quad a^n_2 \sim  a^n_2 + \delta c^n_3 + d c^{n-1}_2
         \end{array}\right\}
          \quad \Longleftrightarrow \quad
        \left\{ \begin{array}{ll}
        a^n_2 = L^a_{\m\n}(x)C_a^{*\m\n}d^n x + \delta b^n_3 + d b_2^{n-1}\,,
\nonumber \\
L^a_{\m\n}(x) = \lambda^a_{\m\n} + A^a_{\m\n\vert\r}x^{\r} + B^a_{\m\n\vert\r\s}x^{\r}x^{\s}\,.
\end{array}\right.
\end{eqnarray}
The constant tensor $\lambda^a_{\m\n}$ is symmetric and traceless
in the indices $\m\n$, and so are the constant tensors
$A^a_{\m\n\vert\r}$ and $B^a_{\m\n\vert\r\s}$. Moreover, the
tensors $A^a_{\m\n\vert\r}$ and $B^a_{\m\n\vert\r\s}$ transform in
the irreducible representations of $GL(n,\mathbb{R})$ labeled
by the Young tableaux
\begin{picture}(22,16)(0,0)
\multiframe(1,4)(10.5,0){2}(10,10){$\m$}{$\n$}
\multiframe(1,-6.5)(10.5,0){1}(10,10){$\r$}
\end{picture}
and
\begin{picture}(25,16)(0,0)
\multiframe(1,4)(10.5,0){2}(10,10){$\m$}{$\n$}
\multiframe(1,-6.5)(10.5,0){2}(10,10){$\r$}{$\s$}
\end{picture}, meaning that
\begin{eqnarray}
        A^a_{\m\n\vert\r} &=& A^a_{\n\m\vert\r}\,,\quad A^a_{(\m\n\vert\r)}\equiv 0\,,
        \nonumber \\
        B^a_{\m\n\vert\r\s} &=& B^a_{\n\m\vert\r\s} = B^a_{\m\n\vert\s\r}\,, \quad
        B^a_{(\m\n\vert\r)\s} = 0\,.
\end{eqnarray}
Together with the tracelessness constraints on the constant
tensors $A^a_{\m\n\vert\r}$ and $B^a_{\m\n\vert\r\s}$, the
$Gl(n,\mathbb{R})$ irreducibility conditions written here above
imply that the tensors $\lambda^a_{\m\n}$, $A^a_{\m\n\vert\r}$ and
$B^a_{\m\n\vert\r\s}$ respectively transform in the
irreducible representations of $O(n-1,1)$ labeled by the Young
tableaux
\begin{picture}(24,16)(0,0)
\multiframe(1,-1)(10.5,0){2}(10,10){$\m$}{$\n$}
\end{picture}, \begin{picture}(22,16)(0,0)
\multiframe(1,4)(10.5,0){2}(10,10){$\m$}{$\n$}
\multiframe(1,-6.5)(10.5,0){1}(10,10){$\r$}
\end{picture}
and
\begin{picture}(25,16)(0,0)
\multiframe(1,4)(10.5,0){2}(10,10){$\m$}{$\n$}
\multiframe(1,-6.5)(10.5,0){2}(10,10){$\r$}{$\s$}
\end{picture}.
\end{prop}
\noindent The proof of Proposition \ref{H2} in the general
spin-$s$ case has been given in Ref. \cite{Bekaert:2005ka} (see also \cite{Barnich:2005bn}).
The spin-$3$ case under consideration was already written in Ref.
\cite{Nazim}.

\subsection{Invariant cohomology of $\d$ modulo $d$: $H_k^{n,inv}(\d \vert\, d)$}
\label{Invariantcharacteristiccohomology}

We have studied above the cohomology of $\delta$ modulo $d$ in the
space of arbitary local functions of the fields $h^a_{\m \n\r}$,
the antifields $\Phi^{*i}$, and their derivatives.  One can also
study $H^n_k(\delta \vert d)$ in the space of invariant
polynomials in these variables, which involve $h^a_{\m\n\r}$ and
its derivatives only through the curvature $K$, the Fronsdal
tensor $F$, and their derivatives (as well as the antifields and
their derivatives). The above theorems remain unchanged in this
space, {\emph{i.e.}} $H_k^{n,inv}(\d \vert\, d)\cong 0$ for $k>2\,$.
This very nontrivial property is crucial for the computation of 
$H^{n,0}(s \vert\, d)$ and is a consequence of
\begin{theorem}\label{2.6}
Assume that the invariant polynomial $a_{k}^{p}$
($p =$ form-degree, $k =$ antifield number) is $\delta$-trivial modulo $d$,
\begin{eqnarray}
a_{k}^{p} = \delta \mu_{k+1}^{p} + d \mu_{k}^{p-1} ~ ~ (k \geqslant 2).
\label{2.37}
\end{eqnarray}
Then, one can always choose $\mu_{k+1}^{p}$ and $\mu_{k}^{p-1}$ to be
invariant.
\end{theorem}
To prove the theorem, we need the following lemma, a proof of which can be found e.g. in
\cite{Boulanger:2000rq}.
\begin{lemma} \label{l2.1}
If $a $ is an invariant polynomial that is $\delta$-exact, $a = \d b$,
then, $a $ is $\delta$-exact in the space of invariant polynomials.
That is, one can take $b$ to be also invariant.
\end{lemma}

The next two subsections are devoted to the proof of Theorem \ref{2.6}.
%
\subsubsection{Propagation of the invariance in form degree}
%
We first derive a chain of
equations with the same structure as (\ref{2.37})
\cite{Barnich:1994mt}. Acting with $d$ on (\ref{2.37}), we get $d
a_{k}^{p} = - \d d \m^{p}_{k+1}$.  Using the lemma and the fact
that $d a_{k}^{p}$ is invariant, we can also write $da_{k}^{p}=
-\d a_{k+1}^{p+1}$ with $a_{k+1}^{p+1}$ invariant. Substituting
this into $d a_{k}^{p} = - \d d \m^{p}_{k+1}$, we get $\d \left[
a_{k+1}^{p+1}-d \m_{k+1}^{p} \right]=0$. As $H(\d)$ is trivial in
antifield number $>0$, this yields \be a_{k+1}^{p+1}=\d
\m^{p+1}_{k+2}+d\m^{p}_{k+1} \ee which has the same structure as
(\ref{2.37}). We can then repeat the same operations, until we
reach form-degree $n$, \be a^{n}_{k+n-p}=\d \m^{n}_{k+n-p+1}+ d
\m^{n-1}_{k+n-p}. \ee

Similarly, one can go down in form-degree. Acting with $\d$ on
(\ref{2.37}), one gets $\d a^{p}_{k}=-d (\d \m^{p-1}_{k})$. If the
antifield number $k-1$ of $\d a^{p}_{k}$ is greater than or equal
to one (i.e., $k>1$), one can rewrite, thanks to Proposition
\ref{2.2}, $\d a^{p}_{k}=-d a^{p-1}_{k-1}$ where $a^{p-1}_{k-1}$
is invariant. (If $k=1$ we cannot go down and the bottom of the
chain is (\ref{2.37}) with $k=1$, namely
$a_1^p=\d\m_2^p+d\m_1^{p-1}$.) Consequently $d \left[
a^{p-1}_{k-1}-\d \m^{p-1}_{k} \right]=0$ and, as before, we deduce
another equation similar to (\ref{2.37}) : \be a^{p-1}_{k-1}=\d
\m^{p-1}_{k}+d\m^{p-1}_{k-1}. \ee Applying $\d$ on this equation
the descent continues. This descent stops at form degree zero or
antifield number one, whichever is reached first, i.e., \bqn
&{\rm{either}}&~~a^{0}_{k-p}=\d \m^{0}_{k-p+1}
\nonumber \\
&{\rm{or}}&~~a^{p-k+1}_{1}=\d \m^{p-k+1}_{2}+d \m^{p-k}_{1}.
\eqn
Putting all these observations together we can write the entire descent as
\bqn
a^{n}_{k+n-p}  &=& \d \m^{n}_{k+n-p+1}+d \m^{n-1}_{k+n-p}
\nonumber \\
& \vdots &
\nonumber \\
a^{p+1}_{k+1}  &=& \d \m^{p+1}_{k+2}+d \m^{p}_{k+1}
\nonumber \\
a^{p}_{k}  &=& \d \m^{p}_{k+1}+d \m^{p-1}_{k}
\nonumber \\
a^{p-1}_{k-1}  &=& \d \m^{p-1}_{k}+d \m^{p-2}_{k-1}
\nonumber \\
& \vdots &
\nonumber \\
{\rm{either}}~~a^{0}_{k-p}&=&\d \m^{0}_{k-p+1}
\nonumber \\
{\rm{or}}~~a^{p-k+1}_{1}&=&\d \m^{p-k+1}_{2}+d \m^{p-k}_{1}
\eqn
where all the $a^{p \pm i}_{k \pm i}$ are invariants.

Let us show that when one of the $\m$'s in
the chain is invariant,
we can actually choose all the other $\m$'s in such a way that they
share this property. In other words, the invariance property propagates up and down in the ladder.
Let us thus assume that $\m^{c-1}_{b}$ is
invariant. This $\m^{c-1}_{b}$ appears in
two equations of the descent :
\bqn
a^{c}_{b} &=& \d \m^{c}_{b+1}+d \m^{c-1}_{b},
\nonumber \\
a^{c-1}_{b-1} &=& \d \m^{c-1}_{b}+ d \m^{c-2}_{b-1} \eqn (if we
are at the bottom or at the top, $\m^{c-1}_{b}$ occurs in only one
equation, and one should just proceed from that one). The first
equation tells us that $ \d \m^{c}_{b+1}$ is invariant. Thanks to
Lemma \ref{l2.1} we can choose $\m^{c}_{b+1}$ to be invariant.
Looking at the second equation, we see that $ d \m^{c-2}_{b-1}$ is
invariant and by virtue of Proposition \ref{2.2}, $\m^{c-2}_{b-1}$
can be chosen to be invariant since the antifield number $b$ is
positive. These two $\m$'s appear each one in two different
equations of the chain, where we can apply the same reasoning. The
invariance property propagates then to all the $\m$'s.
Consequently, it is enough to prove the theorem in form degree
$n$.

\subsubsection{Top form degree}

Two cases may be distinguished depending on whether the antifield number $k$ is greater than $n$ or not.\vspace{.3cm}

In the first case, one can prove the following lemma:
\begin{lemma} \label{l2.2}
If $a^n_k$ is of antifield number $k>n$, then the ``$\m$"s in (\ref{2.37}) can
be taken to be invariant.
\end{lemma}
\noindent{\bf{Proof for $k>n$}} : If $k>n$, the last equation of the
descent is $a^{0}_{k-n}=\d \m^{0}_{k-n+1}$. We can, using Lemma
\ref{l2.1}, choose $\m^{0}_{k-n+1}$ invariant, and so, all the
$\m$'s can be chosen to have the same property.
$\qed$\vspace{.3cm}

It remains therefore to prove Theorem \ref{2.6} in the case where
the antifield number satisfies $k\leqslant n$.
Rewriting the top
equation (i.e. (\ref{2.37}) with $p=n$) in dual notation, we have
\be a_k=\d b_{k+1}+\pa_{\r}j^{\r}_{k},~ (k\geqslant 2).
\label{2.44} \ee We will work by induction on the antifield
number, showing that if the property expressed in Theorem \ref{2.6} is true for $k+1$ (with
$k>1$), then it is true for $k$. As we already know that it is
true in the case $k>n$, the theorem will be proved.\vspace{.4cm}

\noindent{\bf{Inductive proof for $k\leqslant n$}} : The proof follows the lines of
Ref. \cite{Barnich:1994mt} and decomposes in two parts. First, all Euler-Lagrange derivatives of (\ref{2.44}) are computed. Second, the Euler-Lagrange (E.L.) derivative of an invariant quantity is also invariant. This property is used to express the E.L. derivatives of $a_k$ in terms of invariants only. Third, the homotopy formula is used to reconstruct $a_k$ from its E.L. derivatives. This almost ends the proof.\vspace{1.5mm}

{\bf (i)} Let us take the E.L.  derivatives of (\ref{2.44}). Since the E.L.
derivatives with respect to the $C^*_{\a}$ commute with $\d$, we
get first :
\begin{eqnarray}
\frac{\d^L a_k}{\d C^*_{\a\b}} =\d Z^{\a\b}_{k-1}
\label{2.45}
\end{eqnarray}
with
$Z^{\a\b}_{k-1}=\frac{\d^L b_{k+1}}{\d C^*_{\a\b}}$. For the E.L.
derivatives of $b_{k+1}$ with respect to $h^*_{\m\n\r}$ we obtain,
after a direct computation,
\begin{eqnarray}
\frac{\d^L a_k}{\d h^*_{\m\n\r}}=-\d X^{\m\n\r}_k + 3\pa^{(\m}Z^{\n\r)}_{k-1}.
\label{2.46}
\end{eqnarray}
where $X^{\m\n\r}_{k}=\frac{\d^L b_{k+1}}{\d h^*_{\m\n\r} }$.
Finally, let us compute the E.L. derivatives of $a_k$ with respect to the fields.
We get :
\begin{eqnarray}
\frac{\d^L a_k}{\d h_{\m\n\r}}=\d Y^{\m\n\r}_{k+1} + {\cg}^{\m\n\r\vert\a\b\g}
X_{\a\b\g\vert k}
\label{2.47}
\end{eqnarray}
where $Y^{\m\n\r}_{k+1}=\frac{\d^L b_{k+1}}{\d h_{\m\n\r}}$ and
${\cg}^{\m\n\r\vert\a\b\g}(\partial)$ is the second-order self-adjoint differential operator
appearing in the equations of motion (\ref{eom}): $$G^{\m\n\r}={\cg}^{\m\n\r\vert\a\b\g}\,h_{\a\b\g}\,.$$
The hermiticity of $\cg$ implies
${\cg}^{\m\n\r\vert\a\b\g}={\cg}^{\a\b\g\vert\m\n\r}$.\vspace{2mm}

{\bf (ii)} The E.L. derivatives of an invariant object are invariant. Thus,
$\frac{\d^L a_k}{\d C^*_{\a\b}}$ is invariant.  Therefore, by Lemma \ref{l2.1}
and Eq. (\ref{2.45}), we have also
\begin{eqnarray}
\frac{\d^L a_k}{\d C^*_{\a\b}} =\d Z'^{\a\b}_{k-1}
\label{2.45'}
\end{eqnarray}
for some invariant $Z'^{\a\b}_{k-1}$.
Indeed, let us write the decomposition 
$Z^{\a\b}_{k-1} = Z'^{\a\b}_{k-1} + {\tilde{Z}}^{\a\b}_{k-1}$,
where ${\tilde{Z}}^{\a\b}_{k-1}$ is obtained from ${Z}^{\a\b}_{k-1}$ by setting to zero all
the terms that belong only to $H(\g)$. The latter operation clearly commutes
with taking the $\d$ of something, so that Eq. (\ref{2.45}) gives  $0 = \d {\tilde{Z}}^{\a\b}_{k-1}$ which,
by the acyclicity of $\d$, yields ${\tilde{Z}}^{\a\b}_{k-1}=\d \s_k^{\a\b}$ where
$\s_k^{\a\b}$ can be chosen to be traceless.
Substituting $\d \s_k^{\a\b} + Z'^{\a\b}_{k-1}$ for $Z^{\a\b}_{k-1}$ in Eq. (\ref{2.45})
gives Eq. (\ref{2.45'}).

Similarly, one easily verifies that
\begin{eqnarray}
\frac{\d^L a_k}{\d h^*_{\m\n\r}}=-\d X'^{\m\n\r}_k + 3\pa^{(\m}Z'^{\n\r)}_{k-1}\,,
\label{2.46'}
\end{eqnarray}
where $X^{\m\n\r}_k = X'^{\m\n\r}_k + 3\pa^{(\m}\s^{\n\r)}_{k} + \d \r_{k+1}^{\m\n\r}$.
Finally, using ${\cg}^{\m\n\r}{}_{\a\b\g}\,\pa^{(\a}\s^{\b\g)}{}_k=0$ due to
the gauge invariance of the equations of motion ($\s_{\a\b}$ has been taken traceless), we find
\begin{eqnarray}
\frac{\d^L a_k}{\d h_{\m\n\r}} = \d Y'^{\m\n\r}_{k+1}+{\cg}^{\m\n\r}{}_{\a\b\g}
{X'}^{\a\b\g}_k
\label{2.47'}
\end{eqnarray}
for the invariants $X'^{\m\n\r}_k$ and $Y'^{\m\n\r}_{k+1}$.
Before ending the argument by making use of the homotopy formula,
it is necessary to know more about the invariant $Y'^{\m\n\r}_{k+1}$.

Since $a_k$ is invariant, it depends on the fields only
through the curvature $K$, the Fronsdal tensor and their derivatives.
(We remind the reader of  our convention of Section \ref{cohogamma} to substitute $2\pa_{[\g}F_{\r]\m\n}$
for $\eta^{\a\b}K_{\a\m|\b\n|\g\r}$ everywhere.)
We then express the Fronsdal tensor in terms of the
Einstein tensor (\ref{einstein}): $F_{\m\n\r} = G_{\m\n\r} - \frac{3}{n}\eta_{(\m\n}G_{\r)}$, so that we can write $a_k = a_k([\Phi^{*i}],[K],[G])$\,, where $[G]$ denotes the
Einstein tensor and its derivatives.
We can thus write
\begin{eqnarray}
\frac{\d^L a_k}{\d h_{\m\n\r}} = {\cg}^{\m\n\r}{}_{\a\b\g}
{A'}^{\a\b\g}_k + \pa_{\a}\pa_{\b}\pa_{\g}{M'}^{\a\m\vert\b\n\vert\g\r}_k
\label{2.49}
\end{eqnarray}
where $${A'}^{\a\b\g}_k\propto\frac{\d a_k}{\d G_{\a\b\g}}$$ and $${M'}_k^{\a\m\vert\b\n\vert\g\r}\propto {\frac{\d a_k}{\d K_{\a\m\vert\b\n\vert\g\r}}}$$ are both invariant and respectively have the same symmetry properties as the ``Einstein" and ``Riemann" tensors.

Combining Eq. (\ref{2.47'}) with Eq. (\ref{2.49}) gives
\begin{eqnarray}
\d Y'^{\m\n\r}_{k+1} = \pa_\a\pa_\b\pa_\g{M'}_k^{\a\m\vert\b\n\vert\g\r}
                       + {\cg}^{\m\n\r}{}_{\a\b\g} {B'}^{\a\b\g}_k
\label{2.50}
\end{eqnarray}
with ${B'}^{\a\b\g}_k:={A'}^{\a\b\g}_k-{X'}^{\a\b\g}_k$.
Now, only the first term on the right-hand-side of Eq. (\ref{2.50}) is divergence-free,
$\pa_{\m}(\pa_{\a\b\g}{M'}_k^{\a\m\vert\b\n\vert\g\r})\equiv0$,
not the second one which instead obeys
a relation analogous to the Noether identities (\ref{Noether}). As a result, we have
$\d\Big[\pa_{\m}({Y'}^{\m\n\r}_{k+1}-\frac{1}{n}\eta^{\n\r}{Y'}_{k+1}^{\m})\Big]=0\,$,
where ${Y'}_{k+1}^{\m}\equiv \eta_{\n\r}{Y'}_{k+1}^{\m\n\r}\,$.
By Lemma \ref{l2.1}, we deduce
\begin{eqnarray}
        \pa_{\m}({Y'}^{\m\n\r}_{k+1}-\frac{1}{n}\eta^{\n\r}{Y'}_{k+1}^{\m})+\d {F'}_{k+2}^{\n\r}=0
        \,, \label{truc}
\end{eqnarray}
where ${F'}_{k+2}^{\n\r}$ is invariant and can be chosen symmetric and traceless.
Eq. (\ref{truc}) determines a cocycle of $H^{n-1}_{k+1}(d\vert\d)$, for given
 $\n$ and $\r$. Using the general isomorphisms
 $H^{n-1}_{k+1}(d\vert\d)\cong H^{n}_{k+2}(\d\vert d)\cong 0$ ($k\geqslant 1$)
 \cite{Barnich:1994db} gives
\begin{eqnarray}
        {Y'}^{\m\n\r}_{k+1}-\frac{1}{n}\eta^{\n\r}{Y'}_{k+1}^{\m}=
        \pa_{\a}T_{k+1}^{\a\m\vert\n\r} + \delta P^{\m\n\r}_{k+2}
        \,, \label{truc2}
\end{eqnarray}
where both $T^{\a\m\vert\n\r}_{k+1}$ and $P^{\m\n\r}_{k+2}$ are
invariant by the induction hypothesis. Moreover,
$T^{\a\m\vert\n\r}_{k+1}$ is antisymmetric in its first two
indices. The tensors $T^{\a\m\vert\n\r}_{k+1}$ and
$P^{\m\n\r}_{k+2}$ are both symmetric-traceless in $(\n, \r)$.
This results easily from taking the trace of Eq. (\ref{truc2}) with
$\eta_{\n\r}$ and using the general isomorphisms
$H^{n-2}_{k+1}(d\vert\d)\cong H^{n-1}_{k+2}(\d\vert d)\cong
H^{n}_{k+3}(\d\vert d)\cong 0$ \cite{Barnich:1994db} which hold since $k$ is positive. From Eq. (\ref{truc2}) we obtain
\begin{eqnarray}
        {Y'}^{\m\n\r}_{k+1} =
        \pa_{\a} [ T_{k+1}^{\a\m\vert\n\r}+\frac{1}{n-1}\eta^{\n\r}T_{k+1}^{\a\vert\m} ]
        + \delta [ P^{\m\n\r}_{k+2} + \frac{1}{n-1}\eta^{\n\r}P^{\m}_{k+2} ]\,,
         \label{truc3}
\end{eqnarray}
where $T_{k+1}^{\a\vert\m}\equiv \eta_{\n\r}T_{k+1}^{\a\n\vert\r\m}$ and
$P_{k+2}^{\m}\equiv \eta_{\n\r}P_{k+2}^{\n\r\m}\,$.
Since $Y'^{\m\n\r}_{k+1}$ is symmetric in $\m$ and $\n$, we have also
$\pa_{\a}[T_{k+1}^{\a[\m\vert\n]\r}+\frac{1}{n-1}T_{k+1}^{\a\vert[\m}\eta^{\n]\r}]$
$+\;\delta [ P^{[\m\n]\r}_{k+2} + \frac{1}{n-1}\eta^{\r[\n}P^{\m]}_{k+2} ]=0\,$.
The triviality of $H^{n}_{k+2}(d \vert \d)$ ($k>0$) implies again that
$(P^{[\m\n]\r}_{k+2} + \frac{1}{n-1}\eta^{\r[\n}P^{\m]}_{k+2})$ and
$(T_{k+1}^{\a[\m\vert\n]\r}+\frac{1}{n-1}T_{k+1}^{\a\vert[\m}\eta^{\n]\r})$ are trivial,
in particular,
\begin{eqnarray}
T_{k+1}^{\a[\m\vert\n]\r}+\frac{1}{n-1}T_{k+1}^{\a\vert[\m}\eta^{\n]\r}
=\pa_{\b}S^{\b\a\vert\m\n\vert\r}_{k+1}+\d Q^{\a\m\n\r}_{k+2}
\label{derStoT}
\end{eqnarray}
where $S^{\b\a\vert\m\n\vert\r}_{k+1}$ is antisymmetric in ($ \b, \a$) and ($\m,\n$).
Moreover, it is traceless in $\m, \n, \r\,$ as the left hand side of the above equation
shows.
The induction assumption allows us to choose
$S^{\b\a\vert\m\n\vert\r}_{k+1}$ and $Q^{\a\m\n\r}_{k+2}$ invariant.
We now project both sides of Eq. (\ref{derStoT}) on the symmetries of the Weyl tensor.
For example, denoting by $W^{\b\vert\m\n\vert\a\r}_{k+1}$ the projection
${\cw}^{\m\;\n\;\a\;\r\;}_{\m'\n'\a'\r'}S^{\b\a'\vert\m'\n'\vert\r'}_{k+1}$ of
$S^{\b\a\vert\m\n\vert\r}_{k+1}$, we have
\begin{eqnarray}
        W^{\b\vert\m\n\vert\a\r}_{k+1} &=& W^{\b\vert\a\r\vert\m\n}_{k+1}
        = - W^{\b\vert\n\m\vert\a\r}_{k+1} = - W^{\b\vert\m\n\vert\r\a}_{k+1}\,,
        \nonumber \\
        W^{\b\vert\m[\n\vert\a\r]}_{k+1} &=& 0\,,\quad \eta_{\m\a}W^{\b\vert\m\n\vert\a\r}_{k+1}=0\,.
        \nonumber
\end{eqnarray}
As a consequence of the symmetries of $T_{k+1}^{\a\m\vert\n\r}$, the projection of Eq. (\ref{derStoT}) on the symmetries of the Weyl tensor gives
\begin{eqnarray}
        0 = \pa_{\b}W^{\b\vert\m\n\vert\a\r}_{k+1}+ \d (\dots)
\label{derW}
\end{eqnarray}
where we do not write the (invariant) $\d$-exact terms explicitly because they play no role in
what follows.
Eq. (\ref{derW}) determines, for given $(\m, \n, \a, \r)$, a cocycle of
$H^{n-1}_{k+1}(d\vert\d,H(\gamma))$.  Using again the isomorphisms \cite{Barnich:1994db}
$H^{n-1}_{k+1}(d\vert\d)\cong H^{n}_{k+2}(\d\vert d)\cong 0$ ($k\geqslant 1$) and
the induction hypothesis, we find
\begin{eqnarray}
        W^{\b\vert\m\n\vert\a\r}_{k+1} = \pa_{\g}\phi^{\g\b\vert\m\n\vert\a\r}_{k+1} +
        \d (\dots)
        \label{Wintermsofphi}
\end{eqnarray}
where $\phi^{\g\b\vert\m\n\vert\a\r}_{k+1}$ is invariant, antisymmetric in $(\g, \b)$ and
possesses the symmetries of the Weyl tensor in its last four indices. The $\d$-exact
term is invariant as well.
Then, projecting the invariant tensor $4\,\phi^{\g\b\vert\m\n\vert\a\r}_{k+1}$ on the
symmetries of the curvature tensor $K^{\g\b\vert\m\n\vert\a\r}$ and calling the
result $\Psi^{\g\b\vert\m\n\vert\a\r}_{k+1}$ which is of course invariant, we find after some rather lengthy algebra (which takes no time using \emph{Ricci} \cite{Lee})
\be
        {Y'}^{\m\n\r}_{k+1} = \pa_{\a}\pa_{\b}\pa_{\c}\Psi^{\a\m\vert\b\n\vert\c\r}_{k+1}
                              + {\cg}^{\m\n\r}{}_{\a\b\g}\widehat{X}^{\a\b\g}{}_{k+1}+\d(\ldots)\,,
                             \label{result1} \ee
with
\be
        \widehat{X}_{\a\b\g\vert k+1} := \frac{2}{n-2}\cy^{\s\t\r}_{\a\b\c}\Big(
        -S^{\m}_{~\;\s|\m\t|\r\;k+1}
        +\frac{1}{n}\eta_{\s\t}[S_{\m\n|~~\;|\r\;k+1}^{~~~\m\n}+
        S_{\m\n|~\;\r|~~k+1}^{~~~\m~~\n}]\Big)
        \label{result2}
\ee
where $\cy^{\s\t\r}_{\a\b\c}=\cy^{(\s\t\r)}_{(\a\b\c)}$ projects on completely symmetric  rank-3 tensors.\vspace{2mm}

{\bf (iii)} We can now complete the argument. The homotopy formula
\begin{eqnarray}
a_k = \int^{1}_{0}dt\,\left[C^*_{\a\b}\frac{\d^L a_k}{\d C^*_{\a\b}}+
h^*_{\m\n\r}\frac{\d^L a_k}{\d h^*_{\m\n\r}}+
h_{\m\n\r}\frac{\d^L a_k}{\d h_{\m\n\r}}\right](th\,,\,th^*\,,\,tC^*)
\label{homotopy}
\end{eqnarray}
enables one to reconstruct $a_k$ from its E.L. derivatives. Inserting the expressions (\ref{2.45'})-(\ref{2.47'}) for
these E.L. derivatives, we get
\begin{eqnarray}
a_k=\d\Big(\int^{1}_{0}dt\,[C^*_{\a\b}Z'^{\a\b}_{k-1}
+h^*_{\m\n\r}X'^{\m\n\r}_{k}+h_{\m\n\r}Y'^{\m\n\r}_{k+1}](t)\,\Big)+\pa_{\r}k^{\r}.\label{invhf}
\end{eqnarray}
The first two terms in the argument of $\d$ are manifestly invariant. To prove that the third term can be assumed to be invariant in Eq. (\ref{invhf}) without loss of generality, we use Eq. (\ref{result1}) to find that
$$h_{\m\n\r}\,Y'^{\m\n\r}_{k+1}=-\Psi^{\a\m\vert\b\n\vert\c\r}_{k+1}K_{\a\m\vert\b\n\vert\c\r}
+G_{\a\b\g}\widehat{X}^{\a\b\g}{}_{k+1}+\pa_\r \ell^\r+\d(\ldots)\,,$$
where we integrated by part thrice to get the first term of the r.h.s. while
the hermiticity of ${\cg}^{\m\n\r\vert\a\b\g}$ was used to obtain the second term.

We are left with $a_k = \d \m_{k+1} +
\pa_{\r}\n^{\r}_k\,$, where $\m_{k+1}$ is invariant. That
$\n^{\r}_{k}$ can now be chosen invariant is straightforward.
Acting with $\g$ on the last equation yields $\pa_{\r} (\g
{\n}^{\r}_{k}) =0\,$. By the Poincar\'e lemma, $\g {\n}^{\r}_{k} =
\pa_{\s} (\t_k^{[\r \s]})\,$. Furthermore, Proposition \ref{csq}
on $H(\g\vert\, d)$ for positive antighost number $k$ implies that
one can redefine $\n^{\r}_{k}$ by the addition of trivial
$d$-exact terms such that one can assume $\g {\n}^{\r}_{k}=0\,$.
As the pureghost number of ${\n}^{\r}_{k}$ vanishes, the last
equation implies that $\n^{\r}_{k}$ is an invariant
polynomial. \qed

\section{Computation of deformations}
\label{interactions}

As explained in Section \ref{sec:hypotheses}, nontrivial
consistent interactions are in one-to-one correspondance with
elements of $H^{n,0}(s\vert d)$, {\it i.e. }  solutions $a$ of the
equation \be s a+ d b =0\,, \label{topeq}\ee with form-degree $n$
and ghost number zero, modulo the equivalence relation
$$a\sim a+sp+dq\,.$$
Quite generally, one can expand $a$ according to the antifield
number, as \be a=a_0+a_1+a_2+ \ldots a_k\,,\label{antighdec}\ee
where $a_i$ has antifield number $i$.  The expansion stops at some
finite value of the antifield number by locality, as was proved in
\cite{Barnich:1994mt}.

Let us recall \cite{Henneaux:1997bm} the meaning of the various
components of $a$ in this expansion. The antifield-independent
piece $a_0$ is the deformation of the Lagrangian; $a_1$, which is
linear in the antifields $h^{*\m \n\r}$, contains the information
about the deformation of the gauge symmetries, given by the
coefficients of $h^{*\m \n\r}$; $a_2$ contains the information
about the deformation of the gauge algebra (the term $C^{*} C C$
gives the deformation of the structure functions appearing in the
commutator of two gauge transformations, while the term $h^* h^* C
C$ gives the on-shell closure terms); and the $a_k$ ($k>2$) give
the informations about the deformation of the higher order
structure functions and the reducibility conditions.

In fact, using the previous cohomological theorems and standard
reasonings (see e.g. \cite{Boulanger:2000rq}), one can remove all components of $a$ with antifield
number greater than 2. The key point is that the invariant characteristic
cohomology $H^{n,inv}_k(\delta \vert d)$ controls the obstructions to
the removal of the term $a_k$ from $a$ and that all 
$H^{n,inv}_k(\delta \vert d)$ vanish for $k>2$ by Proposition \ref{usefll} and 
Theorem \ref{2.6}. 
This proves the first part of the following theorem:
\begin{theorem}\label{antigh2}Let $a$ be a local top form which is a 
nontrivial solution of the equation (\ref{topeq}). 
Without loss of generality, one can assume
that the decomposition (\ref{antighdec}) stops at antighost number
two, i.e. \be a=a_0+a_1+a_2\,.\label{defdecomps}\ee

If the last term $a_2$ is  parity and Poincar\'e invariant, then
it can always be written as the sum of \be
a_2^2=f^a{}_{bc}\,C^{*\m\n}_{a} (T^b_{\m\a\vert
\b}T^c_{\n\a\vert \b} -2T^b_{\m\a\vert
\b}T^c_{\n\b\vert \a} + \frac{3}{2}\,
C^{b\,\a\b}U^c_{\m\a\vert \n\b})\, d^nx\label{a22}\ee
and \be a^4_2=g^{a}{}_{bc}\,C^{*\m\n}_{a}U^b_{\m\a\vert
\b \l}U^c_{\n\a\vert \b\l}\,d^nx\,,\label{a42}\ee where
$f^a{}_{bc}$ and $g^a{}_{bc}$ are some arbitrary constant tensors
that are antisymmetric under the exchange of $b$ and $c$. 
Notice that $a_2^4$ vanishes when $n=4\,$.
\end{theorem}
\noindent This most general parity and Poincar\'e invariant
expression for $a_2$ is computed in Section \ref{subsec}.

Let us note that the two components of $a_2$ do not contain the
same number of derivatives: $a^2_2$ and $a^4_2$ contain
respectively two and four derivatives. This implies that $a^2_2$
and $a^4_2$ lead to Lagrangian vertices with resp. three and five
derivatives. The first kind of deformation (three derivatives) was
studied in \cite{Berends:1984wp}, however the case with five
derivatives has never been considered before.

Similarly to (\ref{defdecomps}), one can assume $b=b_0+b_1\,.$
Inserting the expansions of $a$ and $b$ into (\ref{topeq}) and
decomposing $s$ as $s=\d+\g$ yields \bqn
\g a_0+\d a_1 +d b_0=0 \,,\label{first}\\
\g a_1+\d a_2 +d b_1=0 \,, \label{second}\\
\g a_2=0\,. \label{minute} \eqn The general solution of
(\ref{minute}) is given by Proposition \ref{Hgamma}. The
computation of $a_2$ follows from the results obtained in Sections
\ref{invPlemma}-\ref{Invariantcharacteristiccohomology}, applied
to the equation (\ref{second}).

Another consequence of the different number of derivatives in
$a^2_2$ and $a^4_2$ is that the descents associated with both
terms can be studied separately. Indeed, the operators appearing
in the descent equations (\ref{first})-(\ref{minute}) are all
homogeneous with respect to the number of derivatives, which means
that one can split $a$ into eigenfunctions of the operator
counting the  number of derivatives and solve the equations
separately for each of them. In the sequel we thus split the
analysis: the descent starting from $a^2_2$ is analysed in Section
\ref{defun}, while the descent associated with  $a^4_2$ is treated
in Section \ref{defdeux}.

\subsection{Most general term in antighost number
two}\label{subsec}

The equation (\ref{minute}) implies that, modulo trivial terms,
$a_2=\a_I \o^I$, where $\a_I $ is an invariant polynomial and the
$\{\o^I\}$ provide a basis of the polynomials in $C_{\m\n},
\widehat{T}_{\m\n\r}, \widehat{U}_{\m\n\r\s}$ (see Section
\ref{cohogamma}). Let us stress that, as $a_2$ has ghost number
zero and antifield number two, $\o^I$ must have ghost number two.

Acting with $\g$ on (\ref{second}) and using the
triviality of $d$, one gets that $b_1$ should also be an element
of $H(\g)$, {\it i.e.}, modulo trivial terms, $b_1=\b_I \o^I$,
where the $\b_I$ are invariant polynomials.

Let us further expand $a_2$ and $b_1$ according to the $D$-degree
defined in the proof of Proposition \ref{csq} in Section
\ref{invPlemma} : \bqn a_2=\sum_{i=0}^M a_2^i= \sum_{i=0}^M
\a_{I_i} \o^{I_i}\,, \hspace{.5cm} b_1=\sum_{i=0}^{M}
b_1^i=\sum_{i=0}^{M} \b_{I_i} \o^{I_i}\,,\nonumber \eqn 
where $a_2^i$, $b_1^i$ and $\o^{I_i}$ have $D$-degree $i$. The
equation (\ref{second}) then reads
$$\sum_i \d [  \a_{I_i} \o^{I_i}]+\sum_i D [\b_{I_i} \o^{I_i}]= \g(\ldots)\,,$$
or equivalently
$$\sum_i \d [  \a_{I_i} ]\o^{I_i}+\sum_i D_0 [\b_{I_i}] \o^{I_i}+\sum_i \b_{I_{i}} A^{I_i}_{I_{i+1}}\o^{I_{i+1}}= \g(\ldots)\,,$$
where $A^{I_i}_{I_{i+1}}\o^{I_{i+1}}= D \o^{I_i}$, which implies
\bqn 
\d [ \a_{I_i} ]+D_0 [\b_{I_i} ]+\b_{I_{i-1}}
A^{I_{i-1}}_{I_{i}}= 0\, \label{third}
\eqn 
for each $D$-degree
$i$, as the elements of the set $\{\o^I\}$ are linearly
independent nontrivial elements of $H(\g)$.
\vspace*{2mm}

\noindent\textbf{\underline{D-degree decomposition}}:

\begin{itemize}
  \item \textbf{degree zero :} In $D$-degree 0, the last equation reads $\d [ \a_{I_0} ]+ D_0
[\b_{I_0} ]=0$, which implies that $\a_{I_0}$ belongs to
$H_2(\d\vert d)$. In antifield number 2, this group has nontrivial
elements given by Proposition \ref{H2}, which are  proportional to
$C^{*\m\n}_{a}$ . The requirement of translation-invariance
restricts the coefficient of $C^{*\m\n}_{a}$  to be constant.
Indeed, it can be shown \cite{book} that if the Lagrangian
deformation $a_0$ is invariant under translations, then so are the
other components of $a$. On the other hand, in $D$-degree 0 and
ghost number 2, we have  $\o^{I_0}=C^b_{\m\r}C^{c}_{\n\s}$. To get
a parity and Lorentz-invariant $a^0_2$, $\o^{I_0}$  must be
completed by multiplication with  $C^{*\m\n}_{a}$ and some
parity-invariant and covariantly constant tensor, {\it i.e.} a
product of $\eta_{\m\n}$'s. The only $a^0_2$ that can be thus
built  is $a_2^0= C^{*\m\n}_{a} C^b_{\m\r}C^{c\r}_{\n}f^{a}_{bc}
d^n x $, where $f^{a}_{bc}$ is some constant tensor that
parametrizes the deformation. From this expression, one computes
that $b^0_1=\b_{I_{0}}\o^{I_0}=-3\, (h^{*\m\n\a}_{a}-\frac{1}{n}\eta^{\m\n} h_a^{*\a})
C^b_{\m\r}C^{c\r}_{\n}f^{a}_{bc}  *  (dx_\a)\,,$ where $*(dx_\a)=\frac{1}{(n-1)!}\e_{\a \m_1 \ldots \m_{n-1}} dx^{\m_{1}}\ldots dx^{\m_{n-1}}$.

  \item \textbf{degree one :} We now analyse Eq.(\ref{third}) in $D$-degree 1, which reads
\be \label{blabla} \d [ \a_{I_1} ]+D_0 [\b_{I_1} ]+\b_{I_{0}}
A^{I_{0}}_{I_{1}}= 0\,.\ee The last term can be read off $
\b_{I_{0}} A^{I_{0}}_{I_{1}}\o^{I_1}\propto
(h^{*\m\n\a}_{a}-\frac{1}{n}\eta^{\m\n} h_a^{*\a}) f^{a}_{bc}
d^{n} x\ \widehat{T}^b_{\a (\m\vert\r)}C^{c\r}_{\n}\,,$  and
should be $\d$-exact modulo $D_0$ for a solution of (\ref{blabla})
to exist. However, the coefficient of $\widehat{T}^b_{\a
(\m\vert\r)}C^{c\r}_{\n}$ is not $\d$-exact modulo $D_0$. This is
easily seen in the space of $x$-independent functions, as both
$\d$ and $D_0$ bring in one derivative while the coefficient
contains none. As $\b_{I_{0}}$ is allowed to depend  explicitely
on $x^\m$, the argument is actually slightly more complicated: one must
expand $\b_{I_{0}}$ according to the number of derivatives of the
fields in order to reach the conclusion. The detailed argument can
be found in the proof of Theorem 7.3 in Ref.
\cite{Henneaux:1996ws}. As $\b_{I_{0}} A^{I_{0}}_{I_{1}}$  is not
$\d$-exact modulo $D_0$, it must vanish if (\ref{blabla}) is to be
satisfied. This implies that $f^{a}_{bc} $ vanishes, so that
$a^0_2=0$ and $b^0_1=0\,.$ One thus gets that $\a_{I_1} $ is an
element of $H_2(\d\vert d)$. However, there is no way to complete
it in a Poincar\'e-invariant way because the only $\o^{I_1}$ is
$\o^{I_1}=\widehat{T}^b_{\m \n\vert\r}C^{c}_{\a \b}$, which has an
odd number of Lorentz indices, while $\a_{I_1} \propto
C^{*\m\n}_{a} $ has an even number of them. Thus $a^1_2=0=b_1^1$.
  \item \textbf{degree two :} The equation (\ref{third}) in $D$-degree 2 is then $\d [ \a_{I_2}
]+D_0 [\b_{I_2} ]=0$, which implies that $\a_{I_2}$ belongs to
$H_2(\d\vert d)$. One finds, most generally when $n>3$, that 
\bqn
a^2_2&= &C^{*\m\n}_{a} (\widehat{T}^b_{\m\a\vert
\b}\widehat{T}_\n^{c\a\vert \b} f^a_{[bc]}+\widehat{T}^b_{\m\a\vert
\b}\widehat{T}_\n^{c\b\vert \a} g^a_{[bc]}+
C^{b\,\a\b}\widehat{U}^c_{\m\a\vert \n\b}k^a_{bc})d^nx\,, \nonumber \\
b_1^2 &= &-3\,(h^{*\m\n\r}_{a}-\frac{1}{n}\eta^{\m\n} h_a^{*\r})
(\widehat{T}^b_{\m\a\vert \b}
\widehat{T}^{c\a\vert\b}_{\n}
f^a_{[bc]}+\widehat{T}^b_{\m\a\vert \b}
\widehat{T}^{c\b\vert\a}_{\n} 
g^a_{[bc]}+ C^{b\,\a\b}\widehat{U}^c_{\m\a\vert
\n\b}k^a_{bc})*(dx_\r)\,, \nonumber \eqn
where $f^a_{[bc]}$,
$g^a_{[bc]}$ and $k^a_{bc}$ are three a priori independent
constant tensors. 
  \item \textbf{degree three :} Now, in the equation for $a^3_2$, we have
  $$\b_{I_{2}} A^{I_{2}}_{I_{3}}\o^{I_3}\propto \Big[ h^{* \m\n\r}_a
\widehat{U}^b_{\m\a\vert \r \b} \widehat{T}_{\n}^{c\a\vert \b}(
f^a_{[bc]}+g^a_{[bc]}-\frac{2}{3}k^a_{cb})-\frac{1}{n}h^{* \r}_a
\widehat{U}^b_{\m\a\vert \r \b} \widehat{T}^{c\m \a\vert \b}(
f^a_{[bc]}+\frac{1}{2}g^a_{[bc]})\Big]\, d^nx\,,
$$
which implies, when $n>3$, that $g^a_{[bc]}=-2 \,f^a_{[bc]}$ and
$k^a_{bc}=\frac{3}{2}\, f^a_{[bc]}\,,$ since the coefficients of $
\widehat{U}^b_{\m\a\vert \r \b} \widehat{T}_{\n}^{c \a\vert \b}$
and $\widehat{U}^b_{\m\a\vert \r \b} \widehat{T}_{\m}^{c \a\vert
\b}  $ are not  $\d$-exact modulo $D_0\,.$ 
All this proves Equation
(\ref{a22}), which is the expression $a_2^2$ found here modulo trivial terms. 
Provided that the above conditions are satisfied,
$\a_{I_3}$ must be in $H_2(\d\vert d)$. But no
Poincar\'e-invariant $a^3_2$ can be built because
$\o^{I_3}=\widehat{T}^b_{\m\a\vert \b}\widehat{U}^c_{\n\r\vert
\s\t}$ has an odd number of Lorentz indices, so $a^3_2=0$.
  \item \textbf{degree four :} 
  Repeating the same arguments for $a^4_2$, one gets
$a^4_2=g^{a}{}_{bc}\,C^{*\m\n}_{a}$
$\widehat{U}^b_{\m\a\vert\b \l}\widehat{U}_\n^{c\a\vert \b\l}d^nx$
and $b^4_1 =-3\,(h^{*\m\n\r}_{a}-\frac{1}{n}\eta^{\m\n}
h_a^{*\r})\widehat{U}^b_{\m\a\vert \b \l}\widehat{U}_\n^{c\a\vert
\b\l} g^{a}_{bc}*\big( dx_\r\big)\,,$ for some constant structure
function $g^{a}_{bc}$.
It is important to notice that $a_2^4$ vanishes in dimension 
$n=4$ because of the Schouten identity $0\equiv C^{*\n_1}_{\m_1} \widehat{U}_{\m_2 \m_3\vert}^{b~~~~\n_2 \n_3} \widehat{U}_{\m_4 \m_5\vert}^{c~~~~\n_4 \n_5}\d^{[\m_1}_{[\n_1} \ldots \d^{\m_5]}_{\n_5]} \propto C^{*\m\n}\widehat{U}^b_{\m\a\vert\b \l}\widehat{U}^{c\a\vert\b\l}_{\n}\,$.
No condition is imposed on $g^{a}_{bc}$ by
equations in higher $D$-degree because $D_1 b^4_1 =0$. 
This proves Equation (\ref{a42}).
  \item \textbf{degree $>4$ :} Finally, there are no $a^i_2$ for $i>4$ because there is no ghost
combination $\o^{I_i}$ of ghost number two and $D$-degree higher
than four.
\end{itemize}
\noindent Summarizing, we have proved the second part of Theorem
\ref{antigh2}.

\subsection{Berends--Burgers--van Dam's deformation}
\label{defun}

In this section, we consider the deformation related to $a^2_2$
given by (\ref{a22}). As explained above, $a_2=a^2_2$  must
now be completed into a solution $a$ of $s a+db=0$ by adding terms
with lower antifield number. The complete solution $a$ provides
then the first-order deformation term $W_1=\int a$ of an
interacting theory. The next step is to check that higher order
terms  $W_2$, $W_3$, etc. can be built to get the full interacting
theory.

In the case considered here, we show that a first-order interaction term $W_1$ can be constructed; however, there is an obstruction to the existence of $W_2$, which prevents its completion into a consistent interacting theory.

\subsubsection{Existence of a first-order deformation}

In this section, the descent equations (\ref{first}) and (\ref{second}), {\it i.e.} $ \g a_0+ \d a_1 +d b_0=0$ and  $\g a_1+ \d a_2 +d b_1=0$,  are solved for $a_1$ and $a_0$.

The latter of these  equations admits the particular solution
\bqn
a_1^p=-\frac{3}{2}\,\Big[\,(h^{*\m\n\r}_{a}-\frac{1}{n}\eta^{\m\n} h_a^{*\r}) \Big(2 \pa_{[\m}h^b_{\a]\b\r}(T^c_{\n\a\vert \b} -2 T^c_{\n\b\vert \a} )+ h^b_{\a\b\r}U^c_{\m\a\vert \n\b}-3 C^{b\,\a\b}\pa_{[\n}h^c_{\b]\r[\a,\m]}\Big) \nonumber \\
+ \frac{1}{n} h^{*\r}_a T^b_{\r\a\vert \b} (\pa_\s h^{c\,
\s\a\b}-\pa^\a h^{c\,\b}-\pa^\b h^{c\,\a})\,\Big]\,f^a_{bc}\,d^nx
\,.\nonumber \eqn To this particular solution, one must add the
general solution $\bar{a}_1$ of $\g \bar{a}_1+d b_1=0\,,$ or
equivalently (by Proposition \ref{csq}) of $\g \bar{a}_1=0$. In
ghost number zero, antifield number one and with two derivatives,
this solution is, modulo trivial $\d$-, $\g$- and $d$-exact terms,
$$\bar{a}_1=h^{*\,a}_{\m\n\r} G^{b\,\m\n}_{\s} C^{c\,\r\s}l^1_{(ab)c}
+h^{*a}_{\m}G^b_{\n}C^{c\,\m\n}l^2_{(ab)c}+h^{*a\,\m}G^b_{\m\n\r}C^{c\,\n\r}l^3_{abc}\,,$$
where $l^1_{(ab)c}$, $l^2_{(ab)c}$ and $l^3_{abc}$ are some arbitrary constants. For future convenience, we also add to $a^p_1+\bar{a}_1$ the trivial term $\g b_1$ where
\bqn
b_1&=& f^a_{bc} h^{*}_{a\m\n\r} (-\frac{3}{2}h^{b\m\s\t}\pa^\n h^{c\r}_{~\s\t}
                                                                -2h^{b\m\s\t}\pa_\s h^{c\n\r}_{~~\t}+3 h^{b \m} \pa^\n h^{c\r}
                                                             -3h^{b}_{\s}\pa^\m h^{c\n\r\s}+2h^{b}_{\s}\pa^\s h^{c\m\n\r})\nonumber \\
&&+f_{abc} h^{*a}_{\m}(2 h^{b\m\n\r}\pa_\n h^{c}_\r -h^{b\m\n\r}\pa^\s h^{c}_{\n\r\s}
                                                        +3 h^{b\m}\pa^\s h^c_{\s}-\frac{1}{2}h^{b}_{\n\r\s}\pa^{\m}h^{c \n\r\s}
                                                        +6 h^{b}_{\n}\pa_\r h^{c\m\n\r})
\,.\nonumber\eqn
In short, up to trivial terms, the most general $a_1$, solution of $\g a_1+ \d a_2 +d b_1=0$, is $a_1=a^p_1+\bar{a}_1+\g b_1\,.$

The next step is to find $a_0$ such that $\g a_0+ \d a_1+d
b_0=0\,.$ A cumbersome but straightforward computation shows that
necessary (and, as we will see, sufficient) conditions for a
solution $a_0$ to exist are (i) $f^a_{[bc]}$ is totally
antisymmetric, or more precisely $\d_{ad}f^d_{[bc]}=f_{[abc]}$,
(ii) $l^1_{(ab)c}=l^2_{(ab)c}=0$ and (iii)
$l^3_{abc}=-\frac{9}{8}f_{[abc]}\,.$ This computation follows the
lines of an argument developped in \cite{Boulanger:2000rq}, which
considers the most general $a_0$ and matches the coefficients of
the terms with the structure $C h'h'$, where $h'$ denotes the
trace of $h$. 
In four dimensions, one must take into account that some of these terms are related by Schouten identities; however, this does not change the conclusions.
Once  the conditions (i) to (iii) are satisfied, one can
explicitly build the solution $a_0$, which  corresponds to the
spin-3 vertex found in \cite{Berends:1984wp} in which the
structure function $f_{abc}$ has been replaced by
$-\frac{3}{8}f_{abc}\,.$ The deformation $a_0$ of the Lagrangian
can be found in the appendix \ref{azeroun}. 
It is unique up to solutions $\bar{a}_0$ of the homogeneous equation 
$\g \bar{a}_0+d b_0=0\,.$

We have thus proved by a new method that the spin-3 vertex of
\cite{Berends:1984wp} is the only consistent nontrivial first-order 
deformation of the free spin-3 theory with at
most\footnote{The developments above prove the three-derivatives
case. For less derivatives, it follows from above that $a_2=0$,
which implies that  $\g a_1= 0$ by (\ref{second}); however there
is no such parity and Poincar\'e-invariant nontrivial $a_1$ with less than
two derivatives, so $a_1=0$ as well.} three derivatives in the
Lagrangian, modulo deformations $\bar{a}_0$ of the latter that are
gauge-invariant up to a total derivative, {\it i.e.} such that $\g
\bar{a}_0+ d b_0=0\,.$ However, as is known from
\cite{Berends:1984rq}, this deformation cannot be completed to all
orders, as is proved again in the next section.

\subsubsection{Obstruction for the second-order deformation}
\label{obstr}

In the previous section, we have constructed a first-order deformation $W_1=\int \,\ll(a_0+a_1+a_2\rr)$ of the free functional $W_0\,.$
As explained in Section \ref{deformation}, a consistent second-order deformation $W_2$ must satisfy the condition \be (W_1,W_1)_{a.b.}=-2 s W_2\,.\label{order2} \ee
Expanding $(W_1,W_1)_{a.b.}$ according to the antifield number, one
finds
$$(W_1,W_1)_{a.b.} = \int d^n x \,(\a_0 + \a_1 + \a_2 )\,,$$
where the term of antifield number two $\a_2$ comes from
the antibracket of $a_2$ with itself.

If one also expands $W_2$ according to the antifield number, one
gets from (\ref{order2}) the following condition on $\a_2$ (it is
easy to see that the expansion of $W_2$ can be assumed to stop at
antifield number three, $W_2 = \int d^n x (c_0 + c_1 + c_2 + c_3)$
and that $c_3$ may be assumed to be invariant, $\g c_3 = 0$) \be
\a_2= -2(\g c_2+\d c_3)+\pa_\m b^\m_2\,. \ee Explicitly, \bqn
\a_2&=\frac{1}{2} f^{}_{abc}f^c_{~de} C^{*a}_{\m\n} &(
-4\widehat{T}^{b \m\a\vert \b}\widehat{T}^{d\n\r\vert
\s}\widehat{U}^e_{\a\r\vert \b\s} +5\widehat{T}^{b \m\a\vert
\b}\widehat{T}^{d\n\r\vert \s}\widehat{U}^e_{\a\s\vert \b\r}
-3\widehat{T}^{b \m\a\vert \b}\widehat{T}^{d}_{\a\r\vert
\s}\widehat{U}_{\hspace{20pt}\b}^{e\s\n\vert \r}
\nonumber \\
&&
+\widehat{T}^{b \m\a\vert \b}\widehat{T}^{d}_{\b\r\vert \s}\widehat{U}_{\hspace{20pt}\a}^{e\r\n\vert \s}
+\widehat{T}^{b \m\a\vert \b}\widehat{T}^{d}_{\b\r\vert \s}\widehat{U}_{\hspace{20pt}\a}^{e\s\n\vert \r}
-\frac{3}{2} \widehat{U}^{b \m\a\vert \n\b}\widehat{T}^d_{\a\r\vert \s}\widehat{T}^{e~\r\vert \s}_{~\b}
\nonumber \\
&&
+3\widehat{U}^{b \m\a\vert \n\b}\widehat{T}^d_{\a\r\vert \s}\widehat{T}^{e~\s\vert \r}_{~\b}
+\frac{9}{4}\widehat{U}^{b \m\a\vert \n\b}C^{d\r\s}\widehat{U}^{e }_{\a\s\vert \b\r}
+\frac{3}{2}C^b_{\a\b}\widehat{U}^{d\r\m\vert \s\a}\widehat{U}^{e~\n\vert \b}_{~\r\hspace{12pt}\s}
\nonumber \\
&&
-\frac{3}{4}C^b_{\a\b}\widehat{U}^{d\r\m\vert \s\a}\widehat{U}^{e~\n\vert \b}_{~\s\hspace{12pt}\r}
+\frac{3}{4}C^{b\a\b}\widehat{U}^{d }_{\r\a\vert \s\b}\widehat{U}^{e\r\m\vert \s\n})+\g(\ldots)\,.
\nonumber
\eqn
It is impossible to get an expression with three ghosts, one $C^{*}$
and no fields, by acting with $\d$ on $c_3$, so we
can assume without loss of generality that $c_3$ vanishes,
which implies that  $\a_2$ should be $\g$--exact modulo total derivatives.

However, $\a_2$ is not a  mod-$d$ $\g$-coboundary unless it vanishes. 
Indeed, suppose we have $$\a_2 = \g(u) +  \pa_\m k^\m  \,.$$ 
Both $u$ and $k^\m$ have antifield number two and we can restrict
ourselves to their components linear in $C^*\,$ without loss of
generality (so that the gauge algebra closes off-shell at second order).
We can also assume that $u$ contains $C^*$
undifferentiated, since derivatives can be removed through
integration by parts. As the Euler derivative of a divergence is
zero, we can reformulate the question as to whether the following
identity holds,
$$ \frac{\d^{L}\a_2 }{\d  C^{* a}_{\m\n}}
 = \frac{\d^{L}(\g u)}{\d  C^{* a}_{\m\n}}=-\g\Big( \frac{\partial^{L}u}{\partial  C^{* a}_{\m\n}}\Big) \,.$$
since $\g C^{*}=0$ and $ C^{* }$ appears undifferentiated in $u$. On the other hand, $
\frac{\d^{L}\a_2 }{\d  C^{* a}_{\m\n}} $ is a sum of nontrivial
elements of $H(\g)$; it can be $\g$-exact only if it vanishes. 
Consequently, a necessary condition
for the closure of the gauge transformations ($c_2$ may be assumed to be linear in the antifields) is $\a_2=0$.

Finally, $\a_2$ vanishes if and only if 
$f_{abc}f^c_{~de} =0\,$ (nilpotency of the algebra) or 
$n=3\,$, which implies when $n>3$ the vanishing of $f_{abc}$ 
(by Lemma \ref{lemalg}), and thus of the whole deformation candidate.

Let us note that originally, in the work \cite{Berends:1984rq}, the
obstruction to this first-order deformation appeared
under the weaker form $f^{}_{abc}f^c_{~de} =f^{}_{adc}f^c_{~be}\,$ 
(associativity) and was obtained by demanding the closure of the algebra of 
gauge transformations at second order in the deformation parameter.

\subsection{Five-derivative deformation}
\label{defdeux}

We now consider the deformation related to $a_2=a^4_2$,
written in Equation (\ref{a42}).
In this case, the general solution $a_1$ of $\g a_1+ \d a_2 +d
b_1=0$ is, modulo trivial terms, 

\be a_1=-2\,(h^{*\m\n\r}_a-\frac{1}{n}
\eta^{\m\n}h^{*\r}_a)\pa^{}_{[\m}h^b_{\a]\r [\b,\l]}
U_\n^{c\a\vert \b\l} g^{a}_{[bc]}\, d^nx +\bar{a}_1 \,,
\label{a1max} \ee
where $\bar{a}_1 $ is an arbitrary element of $H(\g)$.

When the structure constant is completely antisymmetric in its indices, a Lagrangian deformation $a_0$ such that $ \g a_0+ \d a_1 +d b_0=0$ can be computed. However, its expression is quite long and is therefore to be found in the appendix \ref{azerodeux}.
We used the symbolic manipulation program FORM \cite{form} for its computation. 
This nontrivial first-order deformation of the free theory had not been found in the previous spin-three analyzes in Minkowski space-time, which is related to the assumption usually made that the Lagrangian deformation should contain at most three derivatives, while it contains five of them in this case. 

However, it would be very interesting to see whether the cubic vertex  written
in Appendix \ref{azerodeux} could be related to the flat space limit of
the higher-spin vertices of the second reference of \cite{Fradkin:1987ks}.
At first order in the deformation parameter, it might be possible to take
some flat space limit of the $(A)dS_n$ cubic vertices. 
A very appropriate free limit must indeed be taken: The dimensionless coupling
constant $g$ of full higher-spin gauge theory should go to zero in a way that 
compensates the non-analyticity $\sim 1/\Lambda^m$ in the cosmological constant 
$\Lambda$ of the cubic vertices, {\textit{i.e.}} such that the ratio $g/\Lambda^m$
is finite. 
The vertex could then be recovered in such appropriate limits
from the action of \cite{Lopatin:1987hz} by substituting
the linearized spin-$3$ field strengths for the full nonlinear ones at 
quadratic order and replacing the auxiliary and extra connections 
by their expressions in terms of the spin-$3$ gauge field obtained by solving 
the linearized torsion-like constaints, as explained in  
\cite{Vasiliev:2004qz,Sagnotti:2005,Fradkin:1987ks} (and references
therein). 
Such a relation would provide a geometric meaning for the complicated
expression of Appendix \ref{azerodeux}. 

The next step is to find the second order components of the deformation. 
Similarly to the previous case, it can easily be  checked that we can assume $c_3=0$. 
However, no  obstruction arises from the constraint 
$\a_2\equiv (a_2,a_2)=-2 \g c_2+\pa_m k^\m$. 
If this candidate for an interacting theory is obstructed, the obstructions should arise 
at some later stage, {\it i.e.} beyond the (possibly on-shell) closure of the gauge 
transformations.

For completeness, one should check if $ \g a_0+ \d a_1 +d b_0=0$
admits a solution $a_0$ when the structure constant
$g^d_{~bc}=g^d_{~[bc]}$ is not completely antisymmetric but has
the ``hook" symmetry property $\d^{}_{d[a} g^d{}_{bc]}=0$.
However, the computations involved are very cumbersome and we were
not able to reach any conclusion about the existence of such an
$a_0$.

\section{Conclusions and perspectives}
\label{conclusions}

In this paper we carefully analyzed the problem of introducing
consistent interactions among a countable collection of
spin-3 gauge fields in flat space-time of arbitrary dimension
$n> 3\,$. For this purpose we used the powerful BRST
cohomological deformation techniques in order to be as exhaustive
as possible. Under the sole assumptions of locality, parity
invariance, Poincar\'e invariance and perturbative deformation
of the free theory, we proved that only two classes of
non-abelian gauge symmetries are consistent at first order. They
close off-shell and are entirely characterized by the structure
constants of some internal anticommutative algebra (as for
Yang-Mills's theories). When these constant tensors are completely
antisymmetric (this is possible only for a set of different
massless spin-3 fields), there exist actions that are  invariant
at first order under the non-Abelian gauge symmetries. The first
deformation corresponds to the well-known Berends--Burgers--van Dam 
cubic vertex which involves three
derivatives of the fields and becomes inconsistent at second order.  
The second deformation is defined for $n>4$ and corresponds to a cubic vertex
that involves five derivatives. There are some indications that this deformation 
could be obtained from an appropriate flat-space limit of the nonlinear 
$(A)dS_n$ higher-spin gauge theory of Ref. \cite{Vasiliev:2004qz}.

The antisymmetry condition $g_{abc}=g_{[abc]}$ on the structure
constant of the second deformation is only sufficient for the existence of the vertex.
It would be interesting to establish whether a constant tensor
$g^a_{~[bc]}$ with the ``hook" symmetries $\d_{d[a} g^d{}_{bc]}=0$
might not also give rise to a consistent first-order vertex. If
this first-order non-abelian deformation turned out to exist, then
there would be no other one, under the assumptions stated above.
The relaxation of the parity symmetry requirement and the special case $n=3$ 
also deserve more study \cite{Boulanger:2005br}.

Moreover, it would be of prime importance to investigate whether the
second first-order consistent deformation could be extended to higher
orders in the deformation parameter. At second order, a first test
has been passed where the Berends--Burgers--van Dam vertex
fails, but unfortunately the lengthy nature of the five-derivative cubic
vertex makes further analysis very tedious.

Last but not least, it would be of interest to enlarge the set of fields to spin $2$, $3$ and $4$ and see if this allows to remove the previous obstruction at order two. A hint that this might be sufficient comes from the fact that the commutator of two spin-$3$ generators produces spin-$2$ and spin-$4$ generators for the bosonic higher-spin algebra of Ref. \cite{Vasiliev:2004qz}.

\section*{Acknowledgments}
\appendix

N. B. thanks S. Leclercq for discussions.
X. B. acknowledges T. Damour for interesting comments.

\noindent The work of S.C. is supported in part by the ``Interuniversity 
Attraction Poles Programme -- Belgian Science Policy'' and  by 
IISN-Belgium (convention 4.4505.86). Moreover, X.B. and S.C. are 
supported in part by the European Commission FP6 programme 
MRTN-CT-2004-005104, in which S.C. is associated to the V.U.Brussel 
(Belgium).

\section*{Appendices}
\appendix

In this appendix, we provide the Lagrangian deformations $a_0$ for the first-order interactions found in Section \ref{interactions}, as well as the first-order deformation of the gauge transformations for the Berends--Burgers--van Dam vertex.

\section{Three-derivative vertex}
\label{azeroun}

The deformation $$\int a_0\,=\,f_{[abc]}\,S^{abc}\,\,;\quad
S^{abc}[h^d_{\m\n\r}]\,=\, -\frac{3}{8}\,\int{\cal
L}_{BBvD}^{abc}\,d^n x$$ related to the element $a^2_2$ of Section
\ref{defun} is the Berends--Burgers--van Dam cubic vertex \bqn {\cal
L}_{BBvD}^{abc}&=& -\frac{3}{2} \,h^{a \a} h^{b
\b,\,\c}h^c_{\b,\,\a\g} +3\,h^{a \a,\,\b} h^{b \c}h^c_{\g,\,\a\b}
+6 \,h^{a \a\b\g,\,\d} h^{b }_{\a}h^c_{\b,\,\g\d}
+\frac{1}{2}\,h^{a \a} h^{b
\b\c\d,\,\e}h^c_{\b\g\d,\,\a\e}\nonumber \\&& +h^{a
\a}_{~~,\,\a\b} h^{b }_{\g\d\e}h^{c\g\d\e,\,\b} +h^{a \a,\,\b}
h^{b \g\d\e}h^{c}_{\g\d\e,\,\a\b} -3\,h^{a}_{ \a\b\g} h^{b
\a\b}_{~~~\d,\,\e}h^{c \d,\,\g\e} -3\,h^{a}_{ \a\b\g} h^{b
\a\b\d,\,\g\e}h^{c}_{ \d,\,\e}\nonumber \\&& +3\,h^{a}_{
\a\b\g,\,\d} h^{b \a\b\e}h^{c~,\,\g\d}_{ \e} +3\,h^{a~~~,\,\g\d}_{
\a\b\g} h^{b \a\b\e}h^{c}_{ \e,\,\d} -\frac{9}{4}\,h^{a}_{
\a,\,\b\g} h^{b \b}h^{c\g ,\,\a} -\frac{1}{4}\,h^{a}_{ \a,\,\b}
h^{b \b,\,\g}h^{c~,\,\a}_{\g }\nonumber \\&& -3\,h^{a}_{ \a\b\g}
h^{b\d,\,\a}h^{c~,\,\b \g}_{ \d} -\frac{3}{2}\,h^{a ~,\,\a}_{ \a}
h^{b \b,\,\g}h^{c~~~,\,\d}_{\b\g\d} +3\,h^{a}_{ \a}
h^{b}_{\b,\,\g}h^{c~\b\g,\,\a\d}_{ \d} +\frac{3}{2}\,h^{a
~,\,\a\b}_{ \a} h^{b \g,\,\d}h^{c}_{\b\g\d}\nonumber \\&&
+3\,h^{a}_{ \a,\,\b} h^{b}_{\g,\,\d}h^{c\b\g\d,\,\a}
-\frac{3}{2}\,h^{a }_{ \a} h^{b
~~~,\,\b}_{\b\g\d}h^{c~\g\d,\,\a\e}_\e -6\,h^{a
~~~,\,\a\d}_{\a\b\g} h^{b \b,\,\e}h^{c}_{\d\e}{}^\g\nonumber
\\&& +6\,h^{a ~~~,\,\a\d}_{\a\b\g} h^{b \b}h^{c~~\g,\,\e}_{\d\e}
-2\,h^{a}_{ \a\b\g,\,\d} h^{b
~\a\d,\,\e}_{\l}h_\e^{c~\l\b,\g} +h^{a}_{ \a\b\g} h^{b
~~~,\,\a}_{\d\e\l}h^{c~\d\e\l,\,\b\g}\nonumber
\\&& -3\,h^{a}_{ \a\b\g}{}^{,\,\a} h^{b
~\b\g,\,\e}_{\d}h^{c}_{\e\l}{}^{\d,\,\l} +3\,h^{a~~~,\,\a\d}_{
\a\b\g} h^{b \b\g\e,\,\l}h^{c}_{\e\d\l} +6\,h^{a}_{ \a\b\g,\,\d}
h^{b \a\b\e,\,\l}h^{c}_{\e\l}{}^{\d,\,\g}\,, \nonumber \eqn
where we remind that indices after a coma denote partial derivatives.

The first-order deformation of the gauge transformations is given by
 $$\delta^1_{\l}h^a_{\m\n\r} =f^a{}_{bc}\,\Phi^{bc}_{\m\n\r}\,,$$
where $ \Phi^{bc}_{\m\n\r} $ is the completely symmetric component of 
\begin{eqnarray}
        {\cal \phi}^{bc}_{\m\n\r} &= &
6\, h^{b\s}\l^c_{\m\s,\n\r}
-3\, h^{b\s}\l^c_{\m\n,\r\s}
+6\, h^b_{\m,\n}\l^{c\, ~,\s}_{\s\r}
-6\, h^b_{\m}\l^{c~~~\;\s}_{\s\n,\r}
-\frac{15}{4} h^b_{\m\s\t,\n}\l^{c\s,\t}_\r
\nonumber\\ &&
+\frac{31}{4} h^{b\s\t}_{\m}\l^c_{\n\s,\t\r}
+\frac{9}{4} h^b_{\m\n\s,\r\t}\l^{c\s\t}
-\frac{11}{2} h_{\m\n}^{b\; ~\s,\t}\l^c_{\s(\t,\r)}
\nonumber\\ &&-6\, h^b_{\m\n\s,\r}\l^{c\s\t}_{~~~,\t}
-\frac{3}{4} h^b_{\m\s\t,\n}\l^{c\s\t}_{~~~,\r}
-\frac{9}{8} h^b_{\m\s\t,\n\r}\l^{c\s\t}
+\frac{9}{8} h^{b\s\t}_{\m}\l^c_{\s\t,\n\r}
\nonumber\\ &&-\frac{1}{2} h^b_{\m\n\s,\t}\l^{c\t,\s}_\r
+\frac{13}{8} h^{b\s\t}_\m\l^c_{\n\r,\s\t}
+4\, h^b_{\m\n\r,\s}\l^{c\s\t}_{~~~,\t}
-\frac{9}{8} h^{b~~~,\s\t}_{\m\n\r}\l^c_{\s\t}
\nonumber\\
&
+\eta_{\m\n}\Big(&
\frac{9}{4} (h^{b\s,\t}_{~~~\t}\l^c_{\r\s}- h^b_{\s,\r\t}\l^{c\s\t}
- h^{b~~~,\h\s}_{\h\s\t}\l^{c\t}_{\r})
+\frac{9}{8} (h^{b\, ,\s\t}_\s\l^c_{\r\t}+ h^{b\h\s\t}_{~~~~,\h\r}\l^c_{\s\t})
\nonumber\\ &&
+6\,( h^{b\,,\s}_{\s}\l^{c~\,,\t}_{\r\t}- h^{b\s}\l^{c~~~\t}_{\s\t,\r}
- h^{b\s}\l^{c~~~\t}_{\s\r,\t}- h^b_{\s}\l^{c\,~,\s\t}_{\r\t}
- h^b_{\r}\l^{c\,~,\s\t}_{\s\t}
+2\, h^{b~~\,,\s}_{\r\s\t}\l^{c\t,\h}_{\h})
\nonumber\\ &&
+\frac{3}{2} (h^{b\h\s\t}\l^c_{\s\t,\h\r}
-h^b_{\h\s\t,\r}\l^{c\, \s\t,\h})
+ (1-\frac{3}{4n})(2\, h^{b\s,\t}\l^c_{\s\t,\r}-\,h^{b\s\t,\h}_\h \l^c_{\s\t,\r})
\nonumber\\ &&
+(2+\frac{3}{4n})( h^b_{\s,\t}\l_\r^{c\s,\t}+ h^b_{\s,\t}\l_\r^{c\t,\s}
                                        - h_{\s}^{b\t\h,\s}\l^c_{\r\t,\h}- h^b_{\r\s\t}\l_{\h}^{c\s,\t\h}
+ \frac{1}{2}h_\r^{b\s\t}\l^{c~~~\h}_{\s\t,\h})
\nonumber\\ &&
+\frac{9}{8}(1-\frac{1}{n})(- h^{b~~~~\h}_{\r\s\t,\h}\l^{c\s\t}
+2\, h_{\r\s}^{b~\t,\h\s}\l^c_{\h\t}
- h_\r^{b\,,\s\t}\l^c_{\s\t})
\Big)\,.
\nonumber\end{eqnarray}
This expression is equivalent to that of \cite{Berends:1984wp} modulo field redefinitions.

\section{Five-derivative vertex}
\label{azerodeux}

In this appendix, we give the deformation $a_0$ related to the
element $a^4_2$ of Section \ref{defdeux} with completely
antisymmetric structure constants. It satisfies the equation $\g
a_0 + \d a_1+ d b_0=0$ for $a_1$ defined by (\ref{a1max}), in
which $\bar{a}_1=0$. The deformation is
$$ \int a_0\,=\,g^{[abc]}\,T_{abc}\,\,;\quad T_{abc}[h^d_{\m\n\r}]\,=\, \frac{1}{2}\,\int{\cal L}^{}_{abc}\,d^n x$$
where

{\small

\bqn &{\cal L}^{}_{abc}\,=&\nn\\\nn\\ &
h_{a}^{\m\n\r}\,\,\Big(&-{\textstyle \frac{7}{4}}\, \pa_{\m\n}
h_{b}^{\l\s\t}\pa_{\r\s\t}h_{c\l} -{\textstyle \frac{1}{4}}\, \pa_{\m\n}
h_{b}^{\l\s\t}\pa_{\r\h}\pa^\h h_{c\l\s\t} -{\textstyle \frac{1}{2}}\,
\pa_{\m\n} h_{b}^{\l}\pa_{\r\l \s}h_{c}^{\s} -{\textstyle \frac{3}{4}}\,
\pa_{\m\n} h_{b}^{\l}\pa_{\r\s\t}h_{c\l}^{~~\s\t} \nonumber\\&&
-{\textstyle \frac{5}{3}}\, \pa_{\m} h_{b}^{\l\s\t}\pa_{\n\r\l\h}h_{c\s\t}^{\h}
+{\textstyle \frac{1}{2}}\, \pa_{\m} h_{b}^{\l\s\t}\pa_{\n\r\h}\pa^\h
h_{c\l\s\t} +{\textstyle \frac{2}{3}}\, \pa_{\m}
h_{b}^{\l}\pa_{\n\r\s\t}h_{c\l}^{~~\s\t} -{\textstyle \frac{4}{3}}\, \pa_{\m}
h_{b}^{\l}\pa_{\n\r\s}\pa^\s h_{c\l}\nonumber\\&& +{\textstyle \frac{5}{4}}\,
\pa_{\s\t} h_{b}^{\s\t\l}\pa_{\m\n\r} h_{c\l} -{\textstyle \frac{5}{3}}\,
\pa_{\s\t} h_{b}^{\s\l\h}\pa_{\m\n\r} h_{c\l\h}^{\t}
+{\textstyle \frac{3}{4}}\, \pa_{\s}\pa^\s h_{b}^{\l\h\t}\pa_{\m\n\r}
h_{c\l\h\t} +{\textstyle \frac{1}{2}}\, \pa_{\s\t} h_{b}^{\s}\pa_{\m\n\r}
h_{c}^{\t}\nonumber\\&& +{\textstyle \frac{23}{12}}\, \pa_{\s\t}
h_{b}^{\l}\pa_{\m\n\r} h_{c\l}^{~~\s\t} -{\textstyle \frac{4}{3}}\,
\pa_{\s}\pa^{\s} h_{b}^{\l}\pa_{\m\n\r} h_{c\l} -{\textstyle \frac{51}{16}}\,
\pa_{\m\n}h_{b\r}\pa_{\s\t}\pa^\s h_{c}^{\t} -{\textstyle \frac{11}{8}}\,
\pa_{\m} h_{b\n}^{~~\s\t}\pa_{\r\s\t\l} h_{c}^{\l}\nonumber\\&&
+{\textstyle \frac{5}{4}}\, \pa_{\m} h_{b\n \s\t}\pa_{\r\l\h}\pa^\t
h_{c}^{\s\l\h} -{\textstyle \frac{3}{8}}\, \pa_{\m} h_{b\n
\s\t}\pa_{\r\l}\pa^{\l\t} h_{c}^{\s} +{\textstyle \frac{9}{4}}\, \pa_{\m}
h_{b\n\s\t}\pa_{\r\l\h}\pa^\h h_{c}^{\s\t\l} -{\textstyle \frac{1}{12}}\,
\pa_{\m} h_{b\n}\pa_{\r\l \s\t} h_{c}^{\l\s\t}\nonumber\\&&
-{\textstyle \frac{3}{2}}\, \pa_{\m} h_{b\n}\pa_{\r\l \s}\pa^\s h_{c}^{\l}
-{\textstyle \frac{11}{16}}\, \pa_{\l} h_{b\m}^{~\s\t}\pa_{\n\r \s\t}
h_{c}^{\l} -{\textstyle \frac{1}{4}}\, \pa_{\l \h} h_{b\m\s\t}\pa_{\n\r }\pa^\t
h_{c}^{\l\h\s} +{\textstyle \frac{3}{4}}\, \pa_{\l}\pa^{ \l}
h_{b\m\s}^{\t}\pa_{\n\r \t} h_{c}^{\s}\nonumber\\&& +{\textstyle \frac{7}{4}}\,
\pa_{\h\l} h_{b\m}\pa_{\n\r}\pa^\h h_{c}^{~\l} -{\textstyle \frac{19}{16}}\,
\pa_{\h} \pa^\h h_{b\m}\pa_{\n\r\l}h_{c}^{~\l} +{\textstyle \frac{11}{4}}\,
\pa_{\m\l} h_{b\n}^{~~\l\s}\pa_{\s \t\h} h_{c\r}^{~~\t\h}
+{\textstyle \frac{3}{4}}\, \pa_{\m} h_{b\n \s\t}\pa^{\s \t \l\h}
h_{c\r\l\h}\nonumber\\&& +{\textstyle \frac{7}{8}}\, \pa_{\m} h_{b\n
\s\t}\pa^{\s \t \l}\pa_{\l} h_{c\r} +{\textstyle \frac{3}{2}}\, \pa_{\m} h_{b\n
\s\t}\pa^{\s \l}\pa_{\l\h} h_{c\r}^{~~\t\h} -\, \pa_{\m} h_{b\n
\s\t}\pa^{ \l\h}\pa_{\l\h} h_{c\r}^{~~\s\t} +\, \pa_{\m} h_{b\n
}\pa_{ \l}\pa^{\l\s\t} h_{c\r\s\t} \nonumber\\&&+{\textstyle \frac{7}{4}}\,
\pa^{\s} h_{b\m \s\t}\pa^{\t \l\h}\pa_{\n} h_{c\r\l\h}
-{\textstyle \frac{9}{8}}\, \pa^{\s} h_{b\m \s\t}\pa^{\t \l}\pa_{\n\l} h_{c\r}
+{\textstyle \frac{1}{4}}\, \pa^{\l} h_{b\m }^{~~\s\t}\pa_{\n\s\t\h}
h_{c\r\l}^{\h} -{\textstyle \frac{3}{4}}\, \pa^{\l} h_{b\m
}^{~~\s\t}\pa_{\n\s\t\l} h_{c\r}\nonumber\\&& +2\, \pa^{\l\t}
h_{b\m \l\s}\pa_{\n\t\h} h_{c\r}^{~~\s\h} -{\textstyle \frac{1}{4}}\, \pa_{\t}
h_{b\m \l\s}\pa_{\n\h}\pa^{\l\h} h_{c\r}^{~~\s\t} +{\textstyle \frac{3}{4}}\,
\pa^{\t} h^\l_{b\m \s}\pa_{\n\l\t\h} h_{c\r}^{~~\s\h} +\, \pa^{\l}
h_{b\m \s\t}\pa_{\n\l\h}\pa^\h h_{c\r}^{~~\s\t}\nonumber\\&&
-{\textstyle \frac{1}{4}}\, \pa^{\s\t} h_{b\m \s\t}\pa_{\h}\pa^{\h\l}
h_{c\n\r\l} -{\textstyle \frac{3}{4}}\, \pa^{\s} h_{b\m
\s\t}\pa_{\h}\pa^{\t\h\l} h_{c\n\r\l} +{\textstyle \frac{3}{4}}\, \pa^{\l}
h_{b\m \s\t}\pa_{\h}\pa^{\s\t\h} h_{c\n\r\l} \nonumber \\&&
+{\textstyle \frac{3}{2}}\,\pa_{\l} h_{b\m \s\t}\pa^{\l\s\t\h} h_{c\n\r\h}
-{\textstyle \frac{1}{4}}\, \pa^{\l} h_{b\m }\pa_{\s\t}\pa^{\s\t} h_{c\n\r\l}
+{\textstyle \frac{3}{4}}\, \pa^{\l} h_{b\m\l\h }\pa_{\s\t}\pa^{\s\t}
h_{c\n\r}^{\h} +{\textstyle \frac{3}{2}}\, \pa_{\s\t} h_{b\m\l\h }\pa^{\l\s\t}
h_{c\n\r}^{\h}\nonumber \\&&
+{\textstyle \frac{1}{3}}\, \pa_{\m} h_{b\n\r\l }\pa^{\l \s\t\h}h_{c\s\t\h} 
-{\textstyle \frac{15}{4}}\, \pa_{\m} h_{b\n\r\l}\pa^{\l \s\t}\pa_\s h_{c\t} 
-{\textstyle \frac{11}{4}}\, \pa_{\m} h_{b\n\r\l}\pa^{ \s\t\h}\pa_\s h_{c\t\h}^{\l} 
\nonumber\\&&
+{\textstyle \frac{1}{2}}\, \pa_{\m}
h_{b\n\r\l }\pa^{ \s\t}\pa_{\s\t} h_{c}^{\l} +{\textstyle \frac{1}{2}}\,
\pa_{\h} h_{b\m\n\l }\pa^{ \l}\pa_{\r\s\t}
h_{c}^{\h\s\t} 
-{\textstyle \frac{1}{2}}\, \pa_{\h} h_{b\m\n\l}\pa^{ \l\s}\pa_{\r\s} h_{c}^{\h}
\nonumber\\&&
 -\, \pa_{\s} h_{b\m\n\l }\pa^{
\l\s}\pa_{\r\h} h_{c}^{\h} -{\textstyle \frac{3}{4}}\, \pa^{ \h}\pa_{\h}
h_{b\m\n\l }\pa_{\r\s\t} h_{c}^{~\l \s\t} +{\textstyle \frac{1}{2}}\, \pa^{
\s\t} h_{b\m\n\l }\pa_{\r\s\t} h_{c}^{\l }
+{\textstyle \frac{7}{4}}\, \pa^{\l} h_{b\m\n\l }\pa_{\h\s\t}\pa^\h h_{c\r}^{~~\s\t} 
\nonumber\\&&
-{\textstyle \frac{1}{4}}\, \pa^{\l} h_{b\m\n\l }\pa_{\s\t}\pa^{\s\t}
h_{c\r} -{\textstyle \frac{3}{2}}\, \pa^{\h} h_{b\m\n\l }\pa_{\s}\pa^{\l\s\t}
h_{c\r\h\t} -2\,\pa_{\h} h_{b\m\n\l }\pa^{\h\l\s\t}h_{c\r\s\t}
\nonumber\\&& 
+{\textstyle \frac{1}{2}}\, \pa_{\h} h_{b\m\n\l
}\pa^{\h\l\s}\pa_\s h_{c\r} +{\textstyle \frac{1}{4}}\, \pa_{\h} h_{b\m\n\l
}\pa^{\s\t}\pa_{\s\t} h_{c\r}^{~~\h\l} +{\textstyle \frac{1}{2}}\, \pa_{\h}
h_{b\m\n\l }\pa^{\h\s\t}\pa_{\s} h_{c\r\t}^{\l} -{\textstyle \frac{1}{4}}\,
\pa_{\h} h_{b\m\n\r }\pa_{\l\s\t}\pa^{\l} h_{c}^{\h\s
\t}\nonumber\\&& -{\textstyle \frac{3}{8}}\, \pa_{\h} h_{b\m\n\r
}\pa_{\l\s}\pa^{\l\s} h_{c}^{\h} -{\textstyle \frac{1}{2}}\, \pa_{\h}
h_{b\m\n\r }\pa^{\h\l} \pa_{\l\s}h_{c}^{\s} -{\textstyle \frac{27}{16}}\,
\pa_{\m\n} h_{b \l}\pa^{\l\s\t} h_{c\r\s\t} +{\textstyle \frac{15}{16}}\,
\pa_{\m\n} h_{b \l}\pa^{\l\s}\pa_\s h_{c\r}\nonumber\\&&
-{\textstyle \frac{1}{8}}\, \pa_{\m\n} h_{b \l}\pa^{\s}\pa_{\s\h}
h_{c\r}^{~~\l\h} +{\textstyle \frac{1}{3}}\, \pa_{\m} h_{b}^{
\l\s\t}\pa_{\n\l\s\t} h_{c\r} +{\textstyle \frac{1}{2}}\, \pa_{\m\l} h_{b}^{
\l}\pa_{\n\s}\pa^\s h_{c\r} -{\textstyle \frac{33}{16}}\, \pa_{\m} h_{b}^{
\l}\pa_{\n\l\s\t}h_{c\r}^{~~\s\t}\nonumber\\&& -{\textstyle \frac{23}{4}}\,
\pa_{\m}\pa^\s h_{b}^{ \l}\pa_{\n\l\s}h_{c\r} +{\textstyle \frac{5}{8}}\,
\pa_{\m}h_{b}^{ \l}\pa_{\n\s}\pa^{\s\t} h_{c\r\l\t} -3\,
\pa_{\m}h_{b}^{ \l\s\t}\pa_{\n\l\h}\pa^{\h} h_{c\r\s\t}
-{\textstyle \frac{1}{4}}\, \pa_{\l}h_{b}^{ \l\s\t}\pa_{\m\n\s\t}
h_{c\r}\nonumber\\&& -{\textstyle \frac{3}{2}}\,
\pa^{\l\s}h_{b\l}\pa_{\m\n}\pa^\t h_{c\r\s\t} +{\textstyle \frac{11}{4}}\,
\pa^{\l\s}h_{b\l}\pa_{\m\n\s} h_{c\r} -{\textstyle \frac{15}{16}}\,
\pa^{\s\t}h_{b}^{\l}\pa_{\m\n\l} h_{c\r\s\t} +{\textstyle \frac{43}{16}}\,
\pa^{\s}\pa_\s h_{b}^{\l}\pa_{\m\n\l} h_{c\r}\nonumber\\&&
-{\textstyle \frac{11}{4}}\, \pa^{\s\t}h_{b}^{\l}\pa_{\m\n\s} h_{c\r\l\t}
+{\textstyle \frac{19}{8}}\, \pa^{\s}\pa_\s h_{b}^{\l}\pa_{\m\n\t}
h_{c\r\l}^{\t} +{\textstyle \frac{9}{4}}\, \pa_{\h\l}h_{b}^{\h\s\t}\pa_{\m\n\s}
h_{c\r\t}^{\l} +{\textstyle \frac{3}{4}}\, \pa_{\h}h_{b}^{\l\s\t}\pa_{\m\n\s\t}
h_{c\r\l}^{\h}\nonumber\\&& +{\textstyle \frac{15}{4}}\,
\pa_{\l}h_{b}^{\l\s\t}\pa_{\m\n\h}\pa^\h h_{c\r\s\t} -3\,
\pa^{\h}h_{b}^{\l\s\t}\pa_{\m\n\h\l} h_{c\r\s\t} -{\textstyle \frac{1}{2}}\,
\pa_{\m}h_{b\l\s\t}\pa^{\l\s\t\h} h_{c\n\r\h} -{\textstyle \frac{19}{4}}\,
\pa_{\m}h_{b\l}\pa^{\l\s\h}\pa_\s h_{c\n\r\h}\nonumber\\&&
+{\textstyle \frac{1}{2}}\, \pa_{\m}h_{b}^{\l}\pa^{\s\t}\pa_{\s\t} h_{c\n\r\l}
-{\textstyle \frac{5}{2}}\, \pa_{\m}\pa^\h h_{b}^{\l\s\t}\pa_{\h\s\t}
h_{c\n\r\l} -{\textstyle \frac{21}{4}}\, \pa_{\m}h_{b}^{\l\s\t}\pa^\h
\pa_{\h\s\t} h_{c\n\r\l} +{\textstyle \frac{1}{6}}\, \pa_{\l}h_{b}^{\l\s\t}
\pa_{\m\s\t}\pa^\h h_{c\n\r\h}\nonumber\\&& -{\textstyle \frac{1}{2}}\,
\pa^{\h}h_{b}^{\l\s\t} \pa_{\m\l\s\t} h_{c\n\r\h} -5\,
\pa^{\l\h}h_{b\l} \pa_{\m\h\s} h_{c\n\r}^{\s} -{\textstyle \frac{1}{2}}\,
\pa^{\s\h}h_{b}^{\l} \pa_{\m\h\l} h_{c\n\r\s} -{\textstyle \frac{9}{2}}\,
\pa^{\h}\pa_\h h_{b}^{\l} \pa_{\m\l}\pa^\s
h_{c\n\r\s}\nonumber\\&& -{\textstyle \frac{1}{2}}\, \pa^{\s}\pa_\s h_{b}^{\l}
\pa_{\m\h}\pa^\h h_{c\n\r\l} +\, \pa^{\s\t} h_{b}^{\l}
\pa_{\m\s\t} h_{c\n\r\l} -{\textstyle \frac{11}{2}}\, \pa^\h \pa_\s
h_{b}^{\l\s\t} \pa_{\m\t\h} h_{c\n\r\l} +{\textstyle \frac{3}{4}}\, \pa^\h
\pa_\h h_{b}^{\l\s\t} \pa_{\m\s\t} h_{c\n\r\l}\nonumber\\&&
-{\textstyle \frac{1}{4}}\, \pa_\l h_{b}^{\l\s\t} \pa_{\s\t\h}\pa^\h
h_{c\m\n\r} +{\textstyle \frac{1}{2}}\, \pa^\h h_{b}^{\l\s\t} \pa_{\l\s\t\h}
h_{c\m\n\r} -{\textstyle \frac{7}{8}}\,  \pa_\l h_{b}^{\l} \pa_{\s\t}\pa^{\s\t}
h_{c\m\n\r} -{\textstyle \frac{7}{4}}\,  \pa^{\s\t} h_{b}^{\l}
\pa_{\l\s\t}h_{c\m\n\r}\,\Big)\nonumber\\\nn\\&+\,\,h_{a}^{\m}\,\,\Big(&
{\textstyle \frac{1}{2}}\,
 \pa_{\m} h_{b}^{\l\s\t} \pa_{\l\s\t
\r}h_{c}^{\r} -{\textstyle \frac{13}{16}}\,   \pa_{\m} h_{b}^{\s\t\l} \pa_{\s\t
\n\r}h_{c\l}^{~~\n\r} +{\textstyle \frac{9}{16}}\,
 \pa_{\m} h_{b}^{\s\t\l} \pa_{\s\t \n}\pa^\n h_{c\l}
+{\textstyle \frac{1}{2}}\,  \pa_{\m\l}  h_{b}^{\l\n\r} \pa^{\s\t}\pa_\s
h_{c \n\r\t}\nonumber\\&& -{\textstyle \frac{3}{4}}\,  \pa_{\m} h_{b}^{\l\n\r}
\pa_{\l\s} \pa^{\s\t}h_{c \n\r\t} +\,   \pa_{\m} h_{b}^{\l\n\r}
\pa^{\s\t}\pa_{\s\t} h_{c \l\n\r} +{\textstyle \frac{1}{2}}\, \pa_{\m}  h_{b\l}
\pa^{\l\n\r\s} h_{c \n\r\s} -{\textstyle \frac{1}{2}}\,
 \pa_{\m} h_{b}^{\l} \pa_{\l\n\r}\pa^\n h_{c}^{ \r}\nonumber\\&&
+{\textstyle \frac{3}{16}}\,  \pa_{\m}  h_{b}^{\l} \pa^{\n\r\s}\pa_\s
h_{c\l\n\r} +{\textstyle \frac{1}{4}}\,   \pa_{\m} h_{b}^{\l}
\pa^{\r\s}\pa_{\r\s} h_{c\l}-\,  \pa_{\l} h_{b}^{\l\r\s}
\pa_{\m\r\s\t} h_{c}^{\t} +{\textstyle \frac{1}{2}}\, \pa_{\t} h_{b}^{\l\r\s}
\pa_{\m\l\r\s} h_{c}^{\t}\nonumber\\&&  +{\textstyle \frac{23}{16}}\,
 \pa_{\l}  h_{b}^{\l\n\r} \pa_{\m\n\s\t}
h_{c\r}^{~~\s\t} -{\textstyle \frac{3}{4}}\,  \pa_{\l} h_{b}^{\l\n\r}
\pa_{\m\n\s}\pa^\s h_{c\r} -{\textstyle \frac{5}{8}}\,
 \pa^{\l}  h_{b}^{\n\r\s} \pa_{\m\n\r\t} h_{c\l\s}^{\t}
+{\textstyle \frac{25}{4}}\,   \pa_{\l} h_{b}^{\l\n\r} \pa_{\m\s}\pa^{\s\t}
h_{c\n\r\t}\nonumber\\&& +\,  \pa^{\h}  h_{b}^{\l\n\r}
\pa_{\m\l\s}\pa^{\s} h_{c\h\n\r} -6\,  \pa^{\h} h_{b}^{\l\n\r}
\pa_{\m\h\l\s} h_{c\n\r}^\s -\,   \pa^{\h} h_{b}^{\l\n\r}
\pa_{\m\h\s} \pa^{\s} h_{c\l\n\r} -{\textstyle \frac{1}{4}}\,
 \pa^{\l\n} h_{b\l} \pa_{\m\n\r} h_{c}^{\r}\nonumber\\&&
-{\textstyle \frac{1}{2}}\,   \pa_{\n}  h_{b}^{\l} \pa_{\m\l\r\s}
h_{c}^{\n\r\s} +{\textstyle \frac{1}{4}}\,  \pa^{\r\n} h_{b}^{\l}
\pa_{\m\l\r}h_{c\n} -{\textstyle \frac{1}{2}}\,  \pa^{\r}\pa_{\r} h_{b}^{\l}
\pa_{\m\l\n}h_{c}^{\n} +{\textstyle \frac{5}{4}}\, \pa^{\r}\pa_{\r}  h_{b}^{\l}
\pa_{\m\n}\pa^\n h_{c\l}\nonumber\\&& -\,
 \pa^{\n\r}  h_{b}^{\l} \pa_{\m\n\r} h_{c\l}
-{\textstyle \frac{5}{12}}\,   \pa_{\l} h_{b}^{\l\n\r} \pa_{\n\r\s\t}
h_{c\m}^{~~\s\t} +{\textstyle \frac{1}{3}}\,  \pa_{\l} h_{b}^{\l\n\r}
\pa_{\n\r\s}\pa^\s h_{c\m} +{\textstyle \frac{2}{3}}\,
 \pa_{\s} h_{b}^{\l\n\r} \pa_{\l\n\r\t}h_{c\m}^{~~\s\t}\nonumber\\&&
-\,   \pa^{\s} h_{b}^{\l\n\r} \pa_{\l\n\r\s}h_{c\m}
+{\textstyle \frac{9}{16}}\,  \pa_{\l}  h_{b}^{\l} \pa_{\n\r\s}\pa^\n
h_{c\m}^{~~\r\s} +{\textstyle \frac{1}{8}}\,   \pa_{\l}  h_{b}^{\l}
\pa_{\n\r}\pa^{\n\r} h_{c\m} -{\textstyle \frac{3}{8}}\,   \pa_{\n} h_{b}^{\l}
\pa_{\l\r\s}\pa^{\s} h_{c\m}^{~~\n\r}\nonumber\\&& +{\textstyle \frac{3}{8}}\,
 \pa^{\n}  h_{b}^{\l} \pa_{\l\n\r\s} h_{c\m}^{~~\r\s}
-{\textstyle \frac{1}{4}}\,   \pa^{\n}  h_{b}^{\l} \pa^{\r\s} \pa_{\r\s}
h_{c\m\n\l} +{\textstyle \frac{1}{4}}\,   \pa^{\n} h_{b}^{\l} \pa^{\r\s}
\pa_{\n\r} h_{c\m\l\s} -{\textstyle \frac{1}{8}}\, \pa_{\l} h_{b}^{\l\n\r}
\pa^{\s\t} \pa_{\n\s} h_{c\m\r\t}\nonumber\\&& -{\textstyle \frac{3}{4}}\,
 \pa^{\l} h_{b}^{\n\r\s} \pa_{\n\r\t}
\pa^{\t}h_{c\m\l\s} +2\,   \pa^{\l} h_{b}^{\n\r\s} \pa_{\l\n\r\t}
h_{c\m\s}^{\t} -{\textstyle \frac{1}{4}}\,  \pa_{\l} h_{b}^{\l\r\s} \pa^{\n\t}
\pa_{\n\t} h_{c\m\r\s} +{\textstyle \frac{1}{2}}\,
 \pa_{\n} h_{b}^{\l\r\s} \pa^{\n\t} \pa_{\l\t}
h_{c\m\r\s} \nonumber\\&&+{\textstyle \frac{1}{4}}\,  \pa^{\n}  h_{b\m \n\r}
\pa^{\r\l\s\t} h_{c\l\s\t} -{\textstyle \frac{1}{2}}\,   \pa^{\l} h_{b\m \n\r}
\pa^{\n\r\s\t} h_{c\l\s\t} +{\textstyle \frac{3}{16}}\,  \pa^{\l} h_{b\m \n\r}
\pa^{\n\r\s} \pa_\s h_{c\l} -{\textstyle \frac{3}{4}}\,  \pa_{\n} h_{b\m \r\s}
\pa^{\n\r\s \l} h_{c\l}\nonumber\\&& +{\textstyle \frac{9}{4}}\, \pa^{\n\l}
h_{b\m \n\r}  \pa^{\s\t }\pa_\s h_{c\l\t}^{\r} +{\textstyle \frac{3}{2}}\,
 \pa^{\n\l}  h_{b\m \n\r} \pa^{\s\t } \pa_\l
h_{c\s\t}^{\r} +{\textstyle \frac{7}{8}}\,   \pa^{\n}  h_{b\m \n\r} \pa^{\l\s\t
} \pa_\l h_{c\s\t}^{\r} -{\textstyle \frac{1}{2}}\, \pa^{\n} h_{b\m \n\r}
\pa_{\s\t} \pa^{\s\t } h_{c}^{\r}\nonumber\\&& +{\textstyle \frac{1}{2}}\,
 \pa_{\l}  h_{b\m \n\r} \pa_{\s\t} \pa^{\n\t }
h_{c}^{\l\r\s} +{\textstyle \frac{5}{4}}\,  \pa_{\l} h_{b\m \n\r}\pa^{\n\l }
\pa_{\s\t} h_{c}^{\r\s\t} +{\textstyle \frac{1}{2}}\, \pa_{\l} h_{b\m
\n\r}\pa^{\n\l\s } \pa_{\s}   h_{c}^{\r} +{\textstyle \frac{1}{4}}\,
 \pa_{\l}  h_{b\m \n\r}\pa_{\s \t} \pa^{\s\t}
h_{c}^{\l\n\r}\nonumber\\&& -{\textstyle \frac{1}{2}}\,  \pa_{\l} h_{b\m
\n\r}\pa_{\s \t} \pa^{\l\t} h_{c}^{\n\r\s} +{\textstyle \frac{1}{2}}\, \pa_{\l}
h_{b\m }\pa_{\n\r\s } \pa^{\s} h_{c}^{\l\n\r} +{\textstyle \frac{1}{6}}\,
 \pa^{\l}  h_{b\m }\pa_{\l\n\r\s } h_{c}^{\n\r\s}
+{\textstyle \frac{1}{8}}\,   \pa_{\l}  h_{b\m }\pa_{\n\r } \pa^{\n\r }
h_{c}^{\l}\nonumber\\&& -{\textstyle \frac{1}{4}}\,  \pa^{\l} h_{b\m
}\pa_{\l\n\r } \pa^{\n} h_{c}^{\r}\,\Big) \,. \nonumber \eqn

}

\pagebreak

\end{document}